\documentclass[final]{agujournal2019}
\usepackage{url} 
\usepackage{soul}
\draftfalse

\journalname{Space Weather}

\makeatletter
\def\ps@headings{%
  \def\@oddfoot{\centerline{\small --\the\c@page--}}%
  \let\@evenfoot\@oddfoot
  \def\@oddhead{}%
  \let\@evenhead\@oddhead
}
\ps@headings
\makeatother

\begin{document}

%
%


\title{Finding Novel Precursors for Solar Wind Stream Interaction Regions with Interpretable Deep Learning}

%
%




\authors{Prateek Mayank\affil{1,2}, 
Yogesh\affil{3}, 
Enrico Camporeale\affil{1,4}, 
D. Chakrabarty\affil{5},
Lan K Jian\affil{6},
Gregory G. Howes\affil{3},
and Thomas E. Berger\affil{7,1}
}

\affiliation{1}{Space Weather TREC, University of Colorado, Boulder, CO 80303, USA}
\affiliation{2}{University Corporation for Atmospheric Research, CPAESS, Boulder, CO 80303, USA}
\affiliation{3}{Department of Physics and Astronomy, University of Iowa, Iowa City IA 52242, USA}
\affiliation{4}{Queen Mary University of London, London E1 4NS, United Kingdom}
\affiliation{5}{Physical Research Laboratory, Navrangpura, Ahmedabad 380009, India}
\affiliation{6}{Heliophysics Science Division, NASA Goddard Space Flight Center, Greenbelt, MD 20771, USA}
\affiliation{7}{National Center for Atmospheric Research, High Altitude Observatory, Boulder, CO, 80301, USA}








\begin{keypoints}
\item A lightweight Transformer achieves robust per timestep SIR detection with calibrated probability outputs.

\item Feature attribution reveals transverse flow deflection as a robust but previously unquantified SIR signature.

\item The compact interpretable architecture provides a reusable template for deep learning in space weather applications.

\end{keypoints}

%
%

%
%


\vspace{12pt}

\begin{abstract}
Solar wind stream interaction regions (SIRs) drive recurrent geomagnetic storms, yet most existing catalogs rely on expert inspection and simple thresholds that are subjective and can miss events with complex morphologies. We present SIREN (SIR Encoder Network), a lightweight Transformer-based model for per-timestep SIR detection from in-situ solar wind observations. The model ingests sequences of 11 solar wind parameters, spanning magnetic field, velocity, and thermodynamic properties. With approximately 100,000 trainable parameters in a two-layer encoder architecture, SIREN is trained using weighted binary cross-entropy loss and a cosine annealing learning rate. Platt scaling is applied to produce well-calibrated detection probabilities. On a held-out test set of 102 events, the calibrated model achieves a ROC-AUC of 0.93, F1 score of 0.78, and true skill statistic of 0.67. Analysis of the self-attention weights confirms that the model concentrates on the SIR, grounding its decisions in the physically relevant portion of each sequence. Integrated Gradients attribution reveals a quantifiable feature hierarchy: proton density (24.3\%) and magnetic field magnitude (21.6\%) dominate, followed by temperature (13.9\%) and bulk speed (12.1\%). Notably, the transverse velocity component $V_y$ and east-west flow angle together contribute 13–17\%, identifying flow deflection as a consistent but previously under-quantified SIR signature. By producing continuous probabilities rather than binary labels, SIREN enables flexible threshold tuning for operational use and provides a template for compact, interpretable deep-learning systems in space weather.
\end{abstract}

\newpage

\section{Introduction}
Space weather events driven by solar activity pose severe risks to satellites, aviation, communication systems, and electrical infrastructure on Earth. The importance of modeling and forecasting such events has been widely recognized by the scientific community \cite{2015AdSpR..55.2745S, Indian_helio2025, 2026esoar.92354528C} and by stakeholders \cite{ishii2024pathways, national2024next}. Central to space weather forecasting is the solar wind, a continuous structured outflow of plasma from the Sun that modulates heliospheric dynamics and preconditions the near-Earth environment. The solar wind strongly influences the propagation and geoeffectiveness of coronal mass ejections (CMEs) \cite{prateekmayank_2023_swasticme, kay2024} as well as CME-CME interactions \cite{Mayank2024, Smitha2025}. Beyond its role as a modulating medium, the solar wind can also directly drive geomagnetic storms through stream interaction regions (SIRs).

SIRs are large-scale compression structures that form at the interface between fast and slow solar wind streams \cite{Gosling1999, Balogh2013}. When such structures persist over successive solar rotations, they are referred to as corotating interaction regions (CIRs). In contrast to CMEs, which are the dominant sources of intense geomagnetic storms, SIRs are more frequent and typically associated with moderate or recurrent storms that can produce sustained magnetospheric activity \cite{Tsurutani2006, Echer2013, Chi2018}. Statistical studies have shown that approximately one-third to one-half of all SIRs are geoeffective \cite{Alves2006, Zhang2008, Grandin_2019, Hajra_2022}, and the long duration of southward interplanetary magnetic field (IMF) intervals embedded within these structures can drive enhanced magnetospheric convection and substorm activity \cite{Lockwood2016}. SIRs are particularly prevalent during the declining phase and minima of the solar cycle \cite{Jian2011, Grandin_2019}. Given their recurrence, prevalence, and capability for sustained geoeffective driving, reliable identification of SIRs in solar wind data is essential for space weather forecasting.

The plasma and magnetic field parameters within SIRs exhibit characteristic variations as a fast solar wind stream overtakes a slower one. Observations from \textit{Wind} and \textit{ACE} have shown that SIRs are typically marked by a sharp increase in solar wind speed, a pile-up in total pressure, and strong enhancement in proton density and magnetic field strength \cite{Jian2006}. The average duration of SIRs at 1 AU is about 37 hours, with scale sizes of approximately 0.4 AU \cite{Jian2006}. Peak total pressure and magnetic field reach average values of about 176 pPa and 15nT, respectively, while the change in bulk speed between slow and fast streams is roughly 200--250 km s$^{-1}$ \cite{Jian2006}. The helium abundance in the solar wind, which varies depending on its source regions \cite{Yogesh2021, Ofman2024, Yogesh2024}, also evolves during SIRs in response to the interaction between fast and slow streams \cite{Durovcova2019, Yogesh2023}. Although these signatures are well established in the literature, translating them into robust, automated detection criteria remains a challenge.

Despite these well-characterized signatures, most existing SIR catalogs rely on expert inspection of time series of in-situ parameters, where an analyst identifies the onset and offset of stream interaction based on visual assessment of velocity gradients, density and temperature variations, and magnetic field compression with respect to background ambient solar wind \cite{Jian2006, Chi2018}. This process is inherently subjective: two experts examining the same event may place the SIR boundaries at different time steps, particularly at the leading and trailing edges where the transition between ambient wind and compressed plasma is gradual. The resulting catalogs are binary by construction and offer no measure of confidence in the boundary placement. More critically, threshold-based approaches can miss events with complex or atypical morphologies, including weak compressions, extended velocity ramps, or SIRs embedded within compound solar wind structures. An automated data-driven approach that produces calibrated detection probabilities and adapts to diverse event morphologies would address these limitations.

Machine learning (ML) has emerged as a powerful tool for pattern recognition in space weather applications, complementing traditional physics-based and empirical approaches. Deep learning models have demonstrated strong performance in solar flare prediction \cite{Bobra_2015, Vysakh_2023}, solar wind forecasting \cite{Vishal_2020, Mayank_2025}, and geomagnetic index nowcasting \cite{Yuri_2019, Andong_2023}. ML methods have also been applied to the classification of solar wind types \cite{Enrico_2017}, the nowcasting of stream interaction regions \cite{Khaled_enrico_2023} and the detection of CME structures from in-situ measurements \cite{Sanchita_2024, Hannah_2026}. However, most existing approaches focus on event-level classification rather than per-timestep detection, do not produce calibrated probabilities suitable for operational threshold tuning, or lack systematic interpretability analysis that connects model decisions to the underlying physics. For SIR detection specifically, no prior study has combined per-timestep classification with quantitative feature attribution to identify which physical properties the model relies on for detection and how that reliance varies across events.

In this paper, we develop the SIR Encoder Network (SIREN), a lightweight Transformer-based model for per-timestep SIR detection from in-situ solar wind observations and apply systematic feature attribution to understand which physical properties drive the model's decisions. The Transformer's self-attention mechanism is particularly well suited to this task: it enables every timestep to attend to every other timestep in the sequence, allowing the model to capture the multi-day velocity transitions and compression signatures that characterize SIRs. We train and evaluate the model on 676 events from the \citeA{Chi2018} catalog, apply Platt scaling to produce calibrated detection probabilities, and use Integrated Gradients to quantify the contribution of each input feature. Our analysis reveals a quantifiable hierarchy of SIR properties, with proton density and magnetic field magnitude as dominant markers and, notably, transverse flow deflection as a consistent secondary signature, providing new physical insight into the features that define stream interaction regions.

The remainder of the paper is organized as follows. Section 2 describes the data preprocessing, model architecture, training procedure, and feature selection. Section 3 presents the model performance, including training convergence, attention structure, and detection characteristics. Section 4 applies Integrated Gradients to quantify global feature attribution and examines two contrasting case studies. Section 5 validates the learned feature hierarchy against superposed epoch analysis, discusses the physical implications of transverse flow deflection as a diagnostic SIR signature, and addresses limitations. Section 6 summarizes the conclusions.

\section{Methods}

\subsection{Data Preparation} \label{sec: methods}
This study uses in-situ solar wind observations from the Wind spacecraft. Plasma measurements are obtained from the Solar Wind Experiment (SWE) \cite{Ogilvie1995}, and magnetic field observations from the Magnetic Field Investigation (MFI) \cite{Lepping1995}. All datasets are resampled using mean to a uniform cadence of 10 minutes.

The SIR events analyzed in this work are drawn from the catalog compiled by \citeA{Chi2018}, which contains 866 events observed between 1995 and 2016. The catalog provides start, end, and stream interface times for each event; a detailed description of the event selection criteria can be found in \citeA{Chi2018} and references therein. For each event, we select a time window spanning three days before and three days after the stream interface, yielding sequences of 864 timesteps (approximately 6 days). Within each sequence, the SIR interval is labeled with a value of 1 and the surrounding background solar wind with 0. After applying the data quality filtering described in Section 2.1.2, a total of 676 events are retained for the final analysis.

\subsubsection{Feature Selection}
We select 11 solar wind parameters as model inputs, spanning three physically distinct categories relevant to SIR identification.

Magnetic field (4 features): total field magnitude $|B|$ and vector components $B_x$, $B_y$, $B_z$ in Geocentric Solar Ecliptic (GSE) coordinate system. Total field magnitude captures the magnetic compression at the stream interface, while the components encode sector boundary crossings and field deflections associated with the heliospheric current sheet.

Velocity (3 features): bulk speed $V$, and transverse components $V_y$ and $V_z$. Bulk speed captures the defining slow-to-fast velocity transition at SIRs. $V_y$ encodes the east-west flow deflection at the stream interface, which arises from the longitudinal pressure gradient across the compression region. $V_z$ captures any north-south flow perturbation. We exclude $V_x$ from the feature set because it is nearly identical to V (the radial component dominates the bulk speed vector at 1 AU), and including both would introduce a redundant input channel that can dilute gradient-based attribution without adding information.

Thermodynamic and flow properties (4 features): proton density $N_p$, proton temperature $T$, east-west flow angle ($\phi$), and north-south flow angle ($\theta$). Density and temperature are enhanced within the compression region and the post-interface high-speed stream, respectively. The flow angles provide a complementary view of the deflection at the stream interface: $\phi$, in particular, exhibits a characteristic negative excursion as the solar wind is deflected westward ahead of the compression.

Several derived parameters available in the Wind dataset (entropy, plasma beta, Alfvenic Mach number, mass flux, and dynamic pressure) are excluded from the selected feature set. These quantities are computed from the primary measurements listed above and therefore do not carry independent information. Including them would increase the input dimensionality without adding new physical content, while potentially introducing collinear features that complicate gradient-based attribution.

\subsubsection{Data Preprocessing}
To mitigate the influence of outliers, data values below the 0.001 percentile and above the 0.999 percentile are removed for each feature independently. Missing data points are subsequently filled using linear interpolation. To minimize the impact of large data gaps on the analysis, we exclude events in which more than 5\% of the data are missing. After this quality filtering, 676 of the original 866 catalog events are retained.

Prior to normalization, proton temperature is log-transformed to reduce the dynamic range of this inherently skewed distribution. Global min-max normalization is then applied across the full dataset, scaling each of the 11 selected features independently to the [0, 1] interval. To improve stability across different dataset versions, the global minimum and maximum values used for normalization are rounded outward to the nearest order-of-magnitude step (e.g., a raw minimum of 3.7 rounds to 3, a raw maximum of 47.2 rounds to 50). This rounding introduces a negligible loss of resolution but ensures that normalization bounds remain stable if the dataset is extended with additional events.

The dataset is split into training (70\%), validation (15\%), and test (15\%) subsets using a fixed random seed (42) for reproducibility, yielding 473 training events, 101 validation events, and 102 test events.

\subsection{Model}

\begin{figure*}[!ht]
    \centering
    \includegraphics[width = \textwidth]{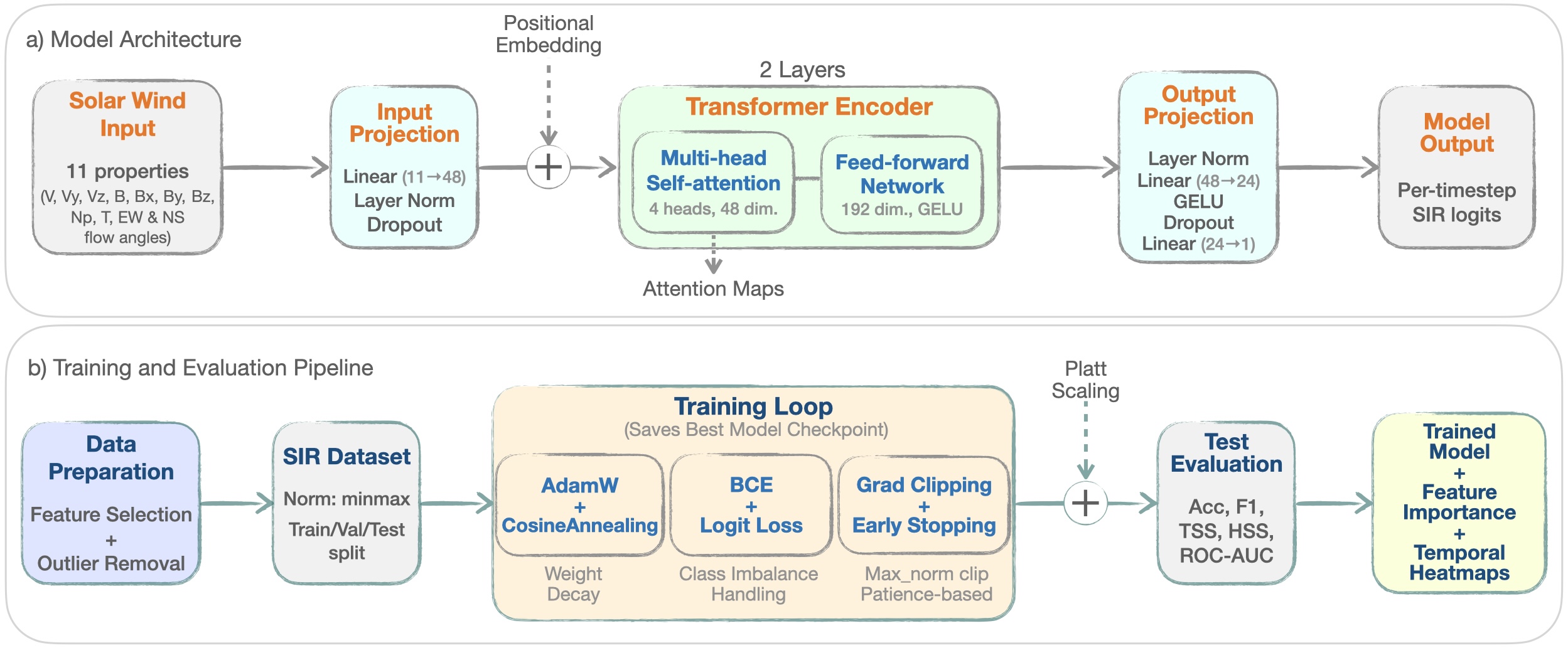}
    \caption{Overview of the SIR detection framework. (a) Model architecture processed by a two-layer Transformer encoder. (b) Training and evaluation pipeline: data preparation, normalization, and splitting feed into a training loop.   
    \label{fig:methodology}}
    \vspace{-1.5cm}
\end{figure*}

We employ a Transformer encoder architecture for per-timestep binary classification of SIR presence in solar wind time series. The model ingests a sequence of 864 timesteps (approximately 6 days at 10-minute cadence) across 11 input features and produces a probability of SIR membership at each timestep.

\subsubsection{Architecture}
The architecture consists of four stages. First, a linear input projection maps the 11-dimensional feature vector at each timestep into a 48-dimensional latent space, followed by layer normalization and dropout. Second, a learned positional embedding of identical dimension is added element-wise to encode temporal order. Third, the projected sequence passes through a two-layer Transformer encoder, where each layer contains a multi-head self-attention block with 4 heads (head dimension = 12) and a position-wise feedforward network with an expansion factor of 4. Both sub-layers use Gaussian Error Linear Unit (GELU) activation, residual connections, layer normalization, and dropout. Fourth, an output head consisting of layer normalization, a two-layer multi-layer perceptron (48 → 24 → 1) with GELU activation and dropout, produces a scalar logit at each timestep.

The complete model contains 99,985 trainable parameters, of which 41,472 reside in the positional embedding and 56,544 in the two Transformer layers. This compact footprint is deliberate: the model is small enough for rapid inference and potential deployment in operational pipelines, while retaining sufficient capacity to capture the multi-scale temporal structure of SIR events.
Self-attention enables every timestep in the sequence to attend to every other timestep, allowing the model to learn long-range dependencies across the full 6-day window. This is particularly relevant for SIR detection, where the velocity transition from slow to fast wind may span several days and the compression signatures at the stream interface are contextualized by upstream and downstream conditions. Unlike recurrent architectures, which process sequences step-by-step, the Transformer processes all timesteps in parallel, enabling both faster training and more direct gradient flow for interpretability analysis.

\subsubsection{Training Procedure}
The model is trained using the AdamW optimizer \cite{AdamW_losh_hutter} with a learning rate of 7 × 10$^{-5}$ and weight decay of 0.02. A cosine annealing schedule reduces the learning rate to a minimum of 10$^{-6}$ over the course of training. Gradients are clipped to a maximum norm of 1.0 to prevent instability.

The loss function is binary cross-entropy with logits, applied independently at each timestep. Because SIR-labeled timesteps constitute a minority of the total sequence (the majority of each 6-day window is ambient wind), we apply a positive class weight of 1.5 to the loss to mitigate class imbalance. This weighting encourages the model to maintain sensitivity to SIR events without excessively penalizing false positives.
Training proceeds for up to 200 epochs with early stopping: if the validation loss does not improve for 20 consecutive epochs, training halts and the model reverts to the best-performing checkpoint.

\subsubsection{Calibration}
Raw sigmoid outputs from a binary classifier are not guaranteed to produce well-calibrated probabilities, particularly when trained with class weighting. To address this, we apply Platt scaling \cite{platt1999} as a post-hoc calibration step. A logistic regression model is fitted on the validation set, mapping raw model logits to calibrated probabilities. This ensures that a predicted probability of, say, 0.8 corresponds to an approximately 80\% empirical frequency of SIR occurrence, which is essential for any operational deployment where users must trade off sensitivity against false-positive rate.

\section{Model Performance}
\begin{figure*}[!ht]
    \centering
    \includegraphics[width = \textwidth]{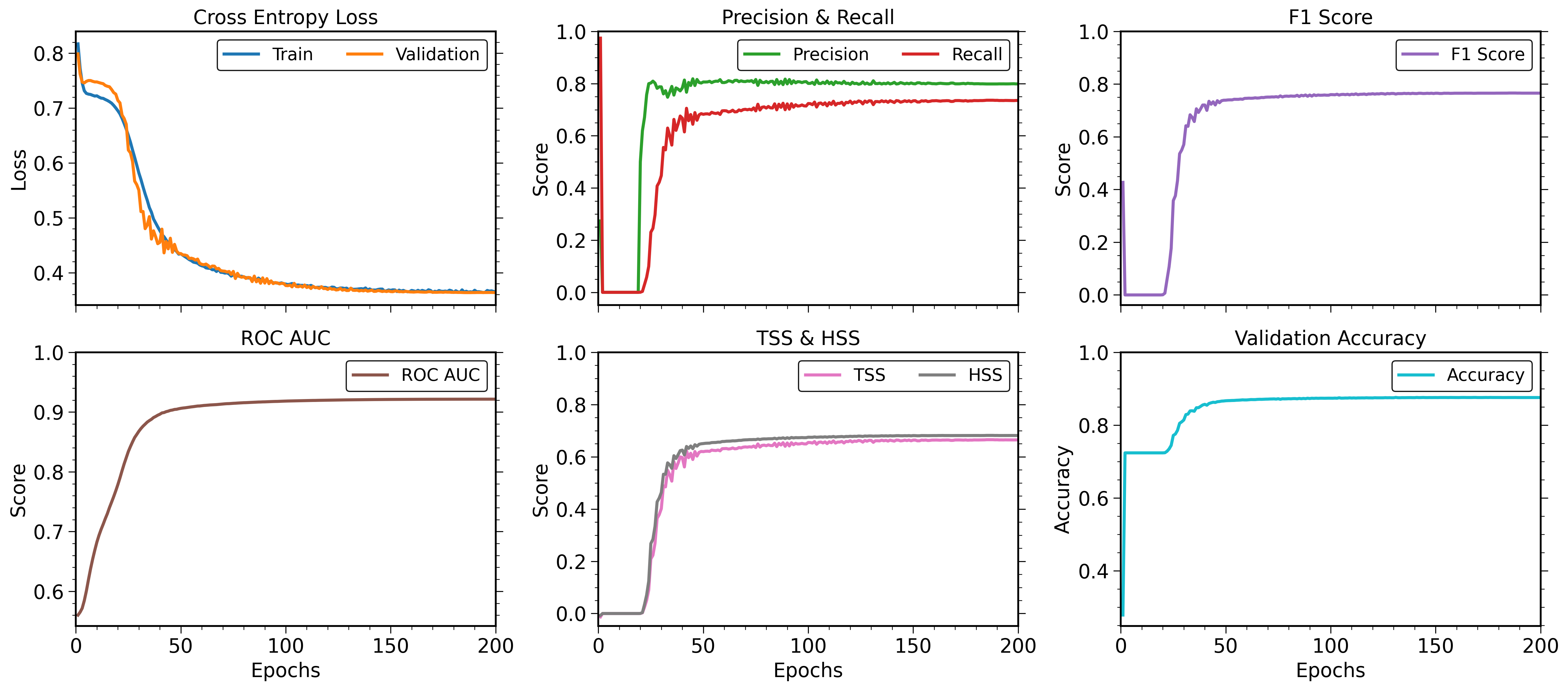}
    \caption{Training history of SIREN over 200 epochs. Top row: training and validation cross-entropy loss (left), precision and recall (center), and F1 score (right). Bottom row: ROC-AUC (left), true skill statistic (TSS) and Heidke skill score (HSS) (center), and validation accuracy (right). All metrics are computed on the validation set except training loss.
    \label{fig:performace_history}}
\end{figure*}

\subsection{Training Convergence and Test-Set Evaluation}

Figure 2 summarizes the training history of the model over 200 epochs using global min–max normalization. The cross-entropy loss decreases steadily for both the training and validation sets during the first 75 epochs, after which both curves plateau and track each other closely through the remainder of training. The absence of a growing gap between training and validation loss indicates that the model does not overfit despite 200 full epochs of training; the combination of dropout (0.4), weight decay (0.02), and the relatively compact architecture (99,985 parameters) provides sufficient regularization for the 676-event dataset. Validation accuracy rises from approximately 0.40 in the first epoch to 0.88 by epoch 75 and remains stable thereafter. The precision and recall curves illustrate the trade-off inherent in the class-weighted loss: precision stabilizes near 0.85, while recall plateaus around 0.72, reflecting the model's conservative detection strategy under the chosen positive class weight of 1.5. The F1 score, which balances precision and recall, converges to approximately 0.78.

The skill scores provide a more informative assessment of detection performance than accuracy alone, given the class imbalance between SIR and non-SIR timesteps. The true skill statistic (TSS) reaches 0.67 and the Heidke skill score (HSS) reaches 0.70 on the validation set, indicating that the model performs substantially better than both random and climatological baselines. The receiver operating characteristic area under the curve (ROC-AUC) stabilizes above 0.90 by epoch 100, confirming strong discrimination between SIR and ambient-wind timesteps across all probability thresholds.


\begin{figure*}
    \centering
    \includegraphics[width = \textwidth]{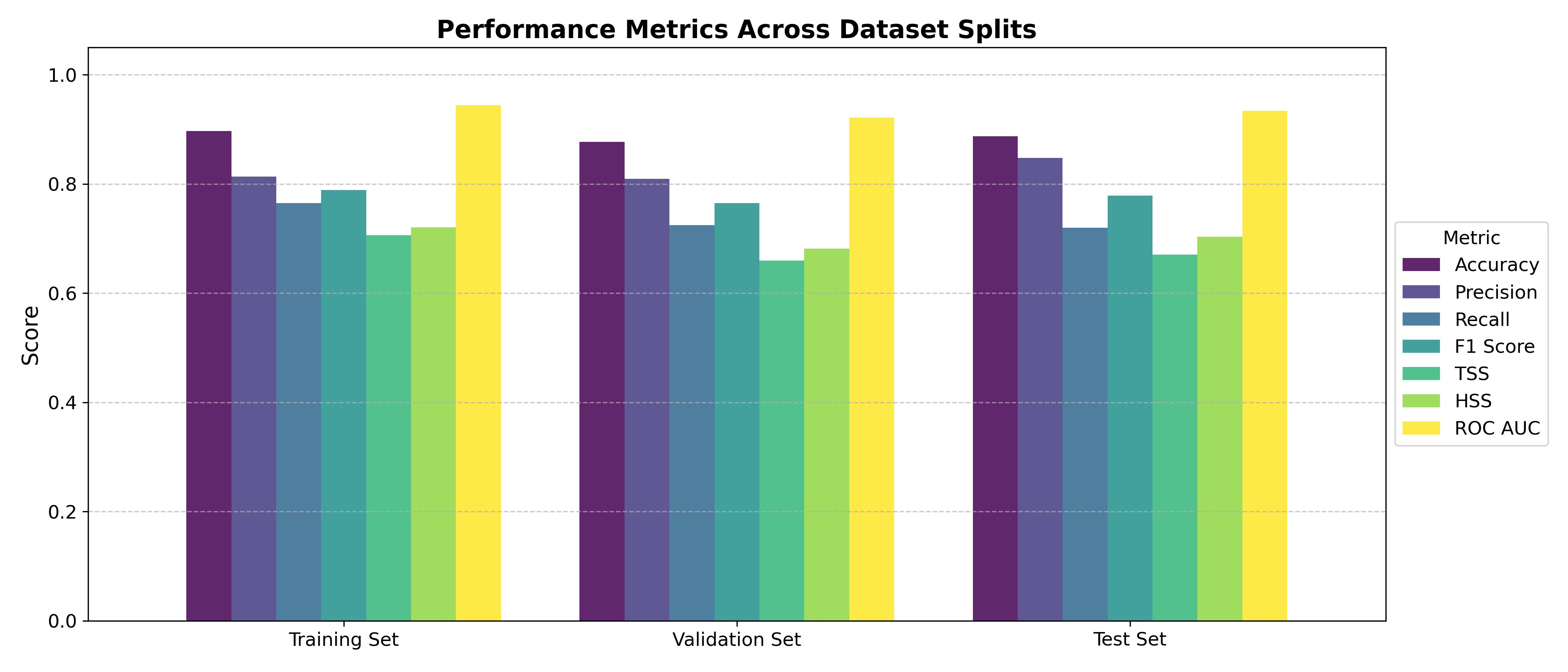}
    \caption{Model performance metrics across training, validation, and test sets.
    \label{fig:allsets_performance}}
    \vspace{-1cm}
\end{figure*}

Figure \ref{fig:allsets_performance} reports the final test-set metrics after Platt calibration, evaluated on 102 held-out events that the model did not encounter during training or validation. The calibrated model achieves an overall accuracy of 88.7\%, with precision of 84.7\% and recall of 72.0\%, yielding an F1 score of 0.78. The TSS of 0.67 and HSS of 0.70 confirm robust skill on unseen data, and the ROC-AUC of 0.93 indicates excellent separability between SIR and non-SIR classes. The modest gap between precision and recall reflects a deliberate design choice: in an operational setting, false positives (flagging ambient wind as SIR) are less disruptive than missed detections, but the class weight of 1.5 prevents the model from becoming excessively permissive at the expense of precision.

\subsection{Attention Structure}

\begin{figure*}[h!]
    \centering
    \includegraphics[width = 0.8\textwidth]{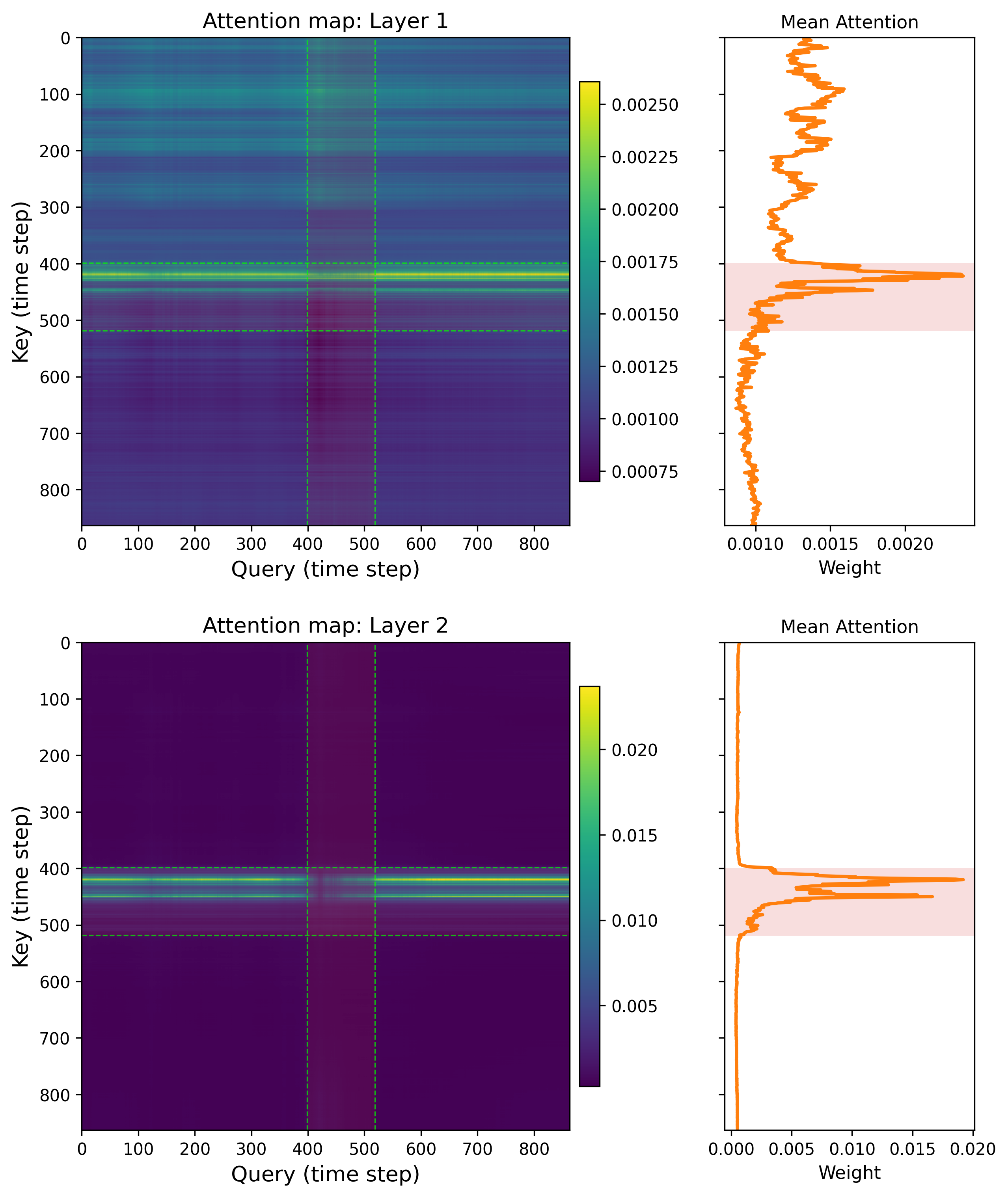}
    \caption{Self-attention weights from the two Transformer encoder layers for a representative test-set sample. Top row is for Layer 1 and bottom row for Layer 2. For each layer, the left panel shows the full attention matrix, with key time steps on the vertical axis and query time steps on the horizontal axis, and the right panel shows the mean attention weight at each key timestep, averaged over all query positions. The shaded band marks the catalog-defined SIR interval.
    \label{fig:attention_map}}
    \vspace{-1cm}
\end{figure*}

To examine what temporal regions the model attends to when making detection decisions, we extract the self-attention weights from both Transformer layers for a representative test-set sample. Figure \ref{fig:attention_map} displays the full attention matrix (left) and the mean attention weight at each key timestep averaged over all query positions (right), for Layer 1 (top) and Layer 2 (bottom). The pink shaded region in the mean attention panels marks the catalog-defined SIR interval.

The two layers exhibit a clear coarse-to-fine progression. In the first layer (Figure \ref{fig:attention_map}, top row), attention is comparatively diffuse. Rather than a single sharp feature, attention in Layer 1 is spread across many time steps, with somewhat higher weight on the upstream slow-wind portion of the sequence. The corresponding mean profile is already raised near the stream interface but retains substantial weight elsewhere, indicating that the first layer aggregates broad temporal context before the representation is refined downstream.

In the second layer (Figure \ref{fig:attention_map}, bottom row), the attention sharpens significantly, and its magnitude in the focused band is roughly an order of magnitude larger than in Layer 1. The matrix exhibits a striking vertical band of elevated attention centered on time steps 400-500, corresponding to the stream interface and surrounding compression region in the 864 timestep sequence. This vertical structure indicates that regardless of where a given query timestep is located (whether in the preceding slow wind, the SIR itself, or the trailing fast wind) the model directs attention toward the same central region where the SIR signatures are concentrated. The associated mean attention profile rises sharply near timestep 380, peaks between time steps 420 and 480, and decays by timestep 550, while attention to the remainder of the sequence is suppressed close to zero.

This temporal localization aligns with the expected position of the SIR within the 6-day window: the catalog stream interface is centered approximately 3 days into each sequence (timestep 432), and the compression region typically spans 1–2 days on either side. Taken together, the two layers show the model first gathering broad temporal context and then sharpening onto the physically relevant compression region. The model has learned to concentrate its representational capacity on the portion of the input sequence where stream interaction signatures occur.

\section{Model Interpretation}
To understand which physical properties drive the model's SIR detection decisions, we apply Integrated Gradients (IG) \cite{Sundararajan_2017}, a gradient-based attribution method that assigns a signed importance value to every input feature at every timestep. The methodology, including the choice of baseline, target formulation, and aggregation procedure, is described in Appendix A.

\subsection{Global Feature Attribution}

\begin{figure*}[h!]
    \includegraphics[width = \columnwidth]{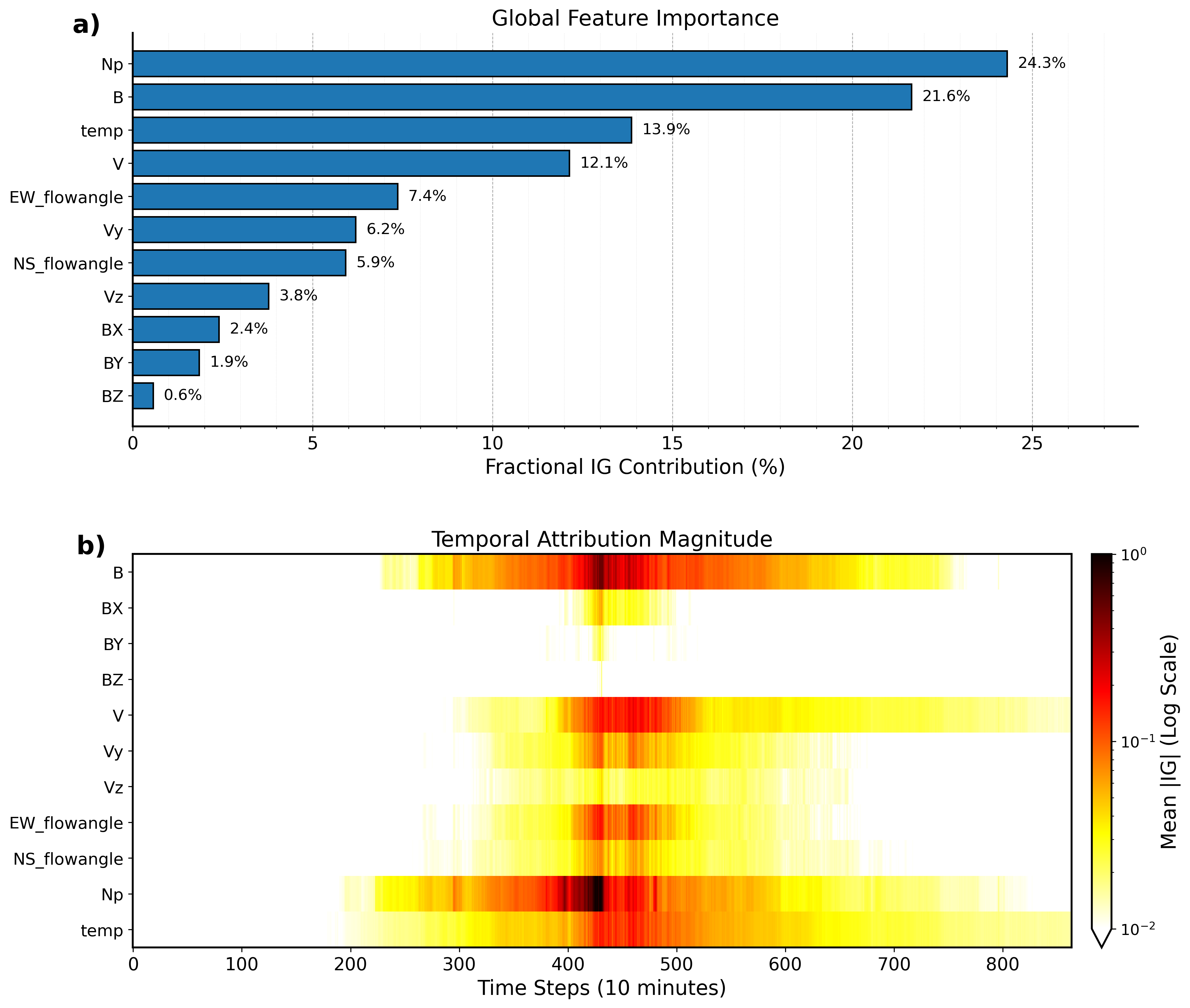}
    \caption{Global feature attribution from Integrated Gradients. (a) Fractional importance of each input feature, computed as mean absolute IG attribution across all 102 test-set samples. (b) Temporal attribution magnitude averaged across all test-set samples, with features on the vertical axis and timesteps on the horizontal axis (log scale).
    \label{fig:ig_summary}}
    \vspace{-1cm}
\end{figure*}

Figure \ref{fig:ig_summary}a presents the global feature importance ranking, computed as the mean absolute IG attribution per feature across all 102 test-set samples and all 864 timesteps per sample. Proton density ($N_p$) emerges as the single most important feature, accounting for 24.3\% of the total attribution. Total magnetic field magnitude $|B|$ ranks second at 21.6\%, followed by temperature (13.9\%) and bulk speed V (12.1\%). Together, these four features account for approximately 72\% of the total attribution, establishing density and magnetic compression as the dominant drivers of the model's SIR detection.

The secondary features form a physically coherent group. The east–west flow angle ($\phi$) contributes 7.4\%, and the transverse velocity component ($V_y$) contributes 6.2\%. These two features encode the same underlying phenomenon, the longitudinal flow deflection at the stream interface, measured in angular and velocity coordinates, respectively. The north–south flow angle (5.9\%) and $V_z$ (3.8\%) provide additional, weaker flow-structure information. The individual magnetic field components $B_x, B_y$, and $B_z$ each contribute less than 3\%, consistent with the physical expectation that total field magnitude, not field direction, is the primary magnetic signature of SIR compression.


Figure \ref{fig:ig_summary}b displays the temporal attribution magnitude averaged across all test-set samples, with features on the vertical axis and timesteps on the horizontal axis. Attribution concentrates sharply at timesteps 350–500, coinciding with the typical SIR center positions in the 6-day window. This temporal localization confirms that the model's feature usage is aligned with the physical location of stream interaction signatures in the input sequence.

The temporal breadth of attribution varies by feature. $N_p$ and $|B|$ exhibit the sharpest and widest temporal localization, with attribution confined primarily to a $\sim$100-timestep window centered on the stream interface. $V$ and temperature, by contrast, show broader temporal spread extending into the post stream interface high-speed stream, consistent with the extended velocity transition and post-interface heating that characterize these features. The flow angles and $V_y$ show intermediate temporal extent, reflecting the deflection signatures that span the compression region.

\subsection{Case Studies}

\begin{figure*}[h!]
    \centering
    \includegraphics[width = 1.1\textwidth]{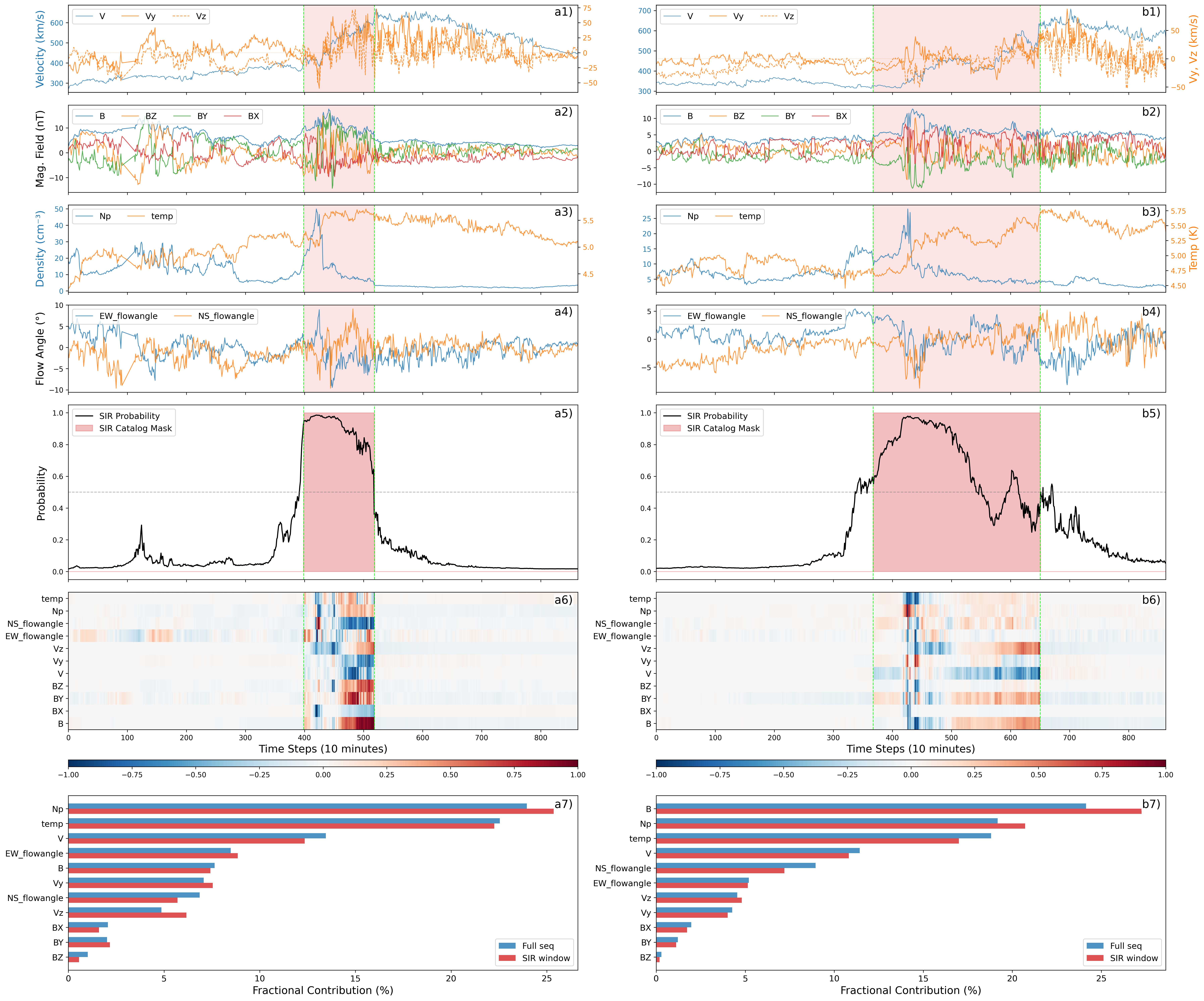}
    \caption{Case studies of two contrasting test-set events. Left column (a1-a7): a strong and well-defined SIR. Right column: a weak and complex SIR. For each event, panels show (top to bottom): solar wind time series (velocity, magnetic field, density and temperature, flow angles), calibrated SIR probability with catalog mask, signed IG attribution heatmap, and per-feature fractional importance for the full sequence (red) and SIR window only (blue).
    \label{fig:sample_comparison}}
    \vspace{-1cm}
\end{figure*}

To illustrate how the model adapts its feature weighting to different event morphologies, we examine two representative test-set events with contrasting SIR signatures. For each event, Figure \ref{fig:sample_comparison} displays the raw solar wind time series, the model's calibrated SIR probability, the signed IG attribution heatmap, and the per-feature importance bar chart.

\subsubsection{Strong and Well-Defined SIR}
The left panels of Figure \ref{fig:sample_comparison} show an event with textbook SIR characteristics: bulk speed increases sharply from approximately 400 km/s to 600 km/s across the stream interface, total magnetic field magnitude rises to $\sim$ 15 nT, and proton density spikes to approximately 40–45 cm$^{-3}$ within the compression region. The temperature rises in the post-interface high-speed stream. The east–west flow angle exhibits a clear negative excursion of 5–10 degrees centered on the stream interface, consistent with the westward deflection of compressed plasma. The model responds with a calibrated probability that reaches about 1.0 within the catalog-defined SIR window, with sharp onset and offset transitions that closely track the catalog boundaries.

The IG attribution for this event is dominated by $N_p$, which accounts for approximately 22–25\% of the total importance, consistent with its global ranking as the single most important feature. Temperature ranks second, followed by the bulk speed $V$ and east–west flow angle, while $|B|$ contributes a more moderate share despite the clear magnetic compression visible in the time series. The signed heatmap shows concentrated attribution within the SIR window, with the characteristic alternating positive and negative pattern for $N_p$ reflecting the density fluctuations discussed above. The east–west flow angle shows a coherent band of attribution centered on the stream interface, highlighting the model's sensitivity to the deflection signature in this well-defined event. This case exemplifies the regime where density compression and flow deflection together dominate the detection decision, and where the model's probability output closely reproduces the catalog boundaries.

\subsubsection{Weak and Complex SIR}
The right panels of Figure \ref{fig:sample_comparison} show a contrasting event: a more gradual velocity transition, moderate magnetic field enhancement, and subdued density compression compared to the strong-SIR case. Despite these attenuated signatures, the model still achieves elevated SIR probability within the catalog window, though with a more gradual onset and noticeable mid-event modulation where the probability temporarily dips before recovering. These probability fluctuations correspond to intervals where the local plasma conditions momentarily depart from the compression signature, demonstrating the model's continuous evaluation of input features rather than a simple on/off detection.

The feature importance profile for this event differs from the strong-SIR case. Total magnetic field magnitude $|B|$ emerges as the leading contributor, displacing $N_p$ from its dominant position in the strong-SIR event. This inversion reflects the model's adaptive strategy: in an event where the density enhancement is modest, the model shifts its reliance toward the magnetic compression that remains the most distinctive local signature. The north–south flow angle and east–west flow angle both gain relative importance compared to their global averages, rising to approximately 5–7\% each. This elevated contribution from flow structure is consistent with the physical expectation that transverse deflection signatures persist even when the primary compression indicators are subdued. The transverse velocity $V_y$ also contributes, reinforcing the finding that flow deflection provides complementary diagnostic value across a range of event strengths.

\subsubsection{Interpreting Negative Density Attribution}
Although $N_p$ ranks first in absolute importance, its signed attribution reveals more complex behavior. The signed temporal heatmap shows that $N_p$ has net negative attribution at several timesteps within the SIR center, despite being the highest-ranked feature by magnitude. This seemingly paradoxical result reflects the physical dynamics of density within the interaction region.

Proton density exhibits rapid fluctuations within the SIR: sharp enhancements at compression fronts are interspersed with local dips where the plasma momentarily rarefies. Each density increase relative to the pre-event baseline generates positive IG attribution (pushing the model toward SIR detection), while each local dip generates negative attribution (pushing the model away from detection). When averaged across samples, the dips slightly dominate at certain timesteps, producing net negative signed attribution in the aggregate. The magnitude-based importance correctly captures density's overall diagnostic role regardless of sign.

This sensitivity to density's temporal fine structure, rather than to its mean level, is a notable property of the model. A simple threshold-based detector would flag density above a fixed value as indicative of SIR presence; the Transformer, by contrast, has learned to track the temporal dynamics of compression, distinguishing between sustained density enhancements (strong SIR evidence) and transient fluctuations (weaker or ambiguous evidence). This capacity to evaluate feature dynamics rather than static values is a direct consequence of the self-attention mechanism, which allows the model to contextualize each timestep within the full sequence.


\subsection{Real-Time Detection Capabilities}

\begin{figure*}[h!]
    \centering
    \includegraphics[width = \textwidth]{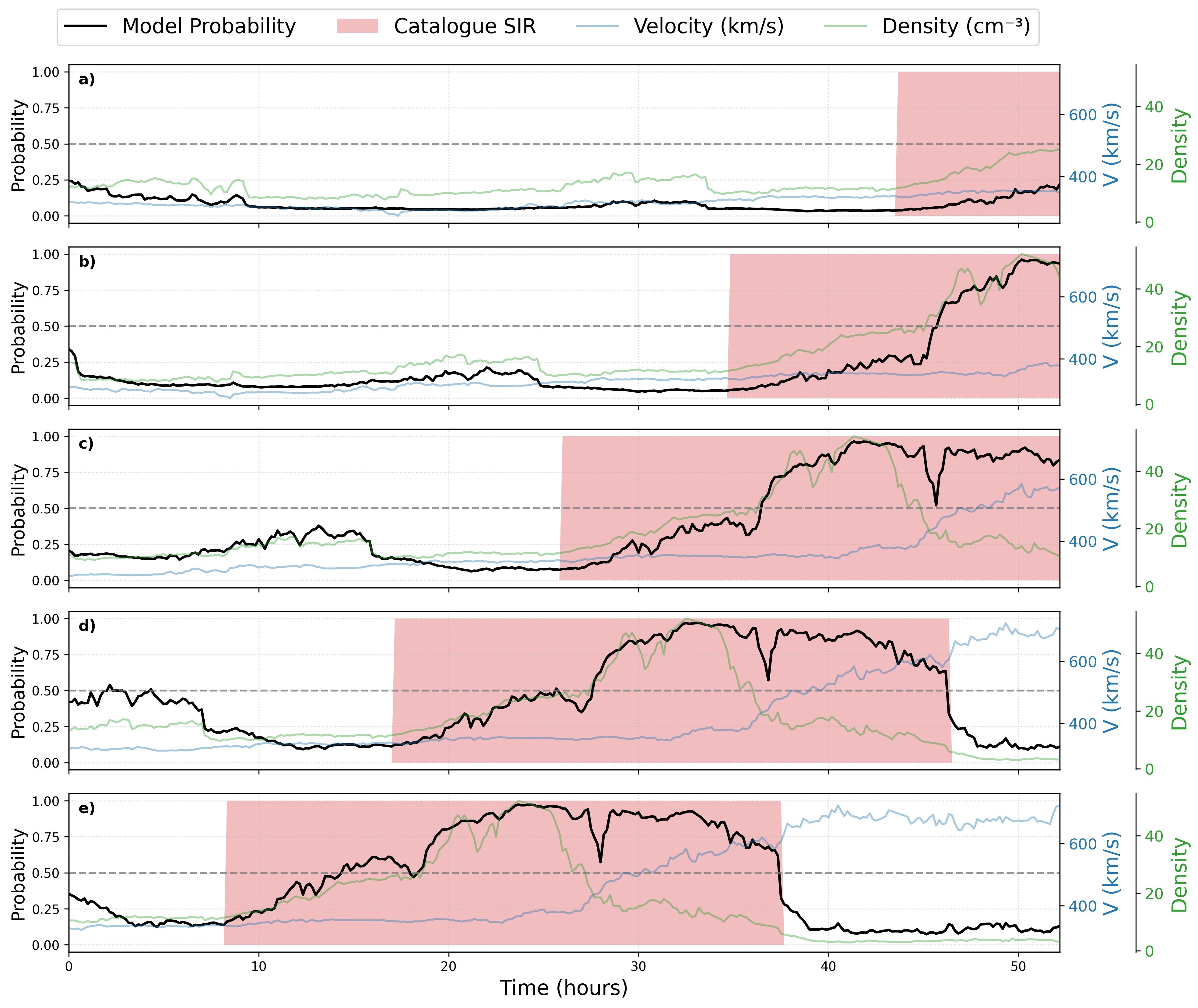}
    \caption{Real-time response of SIREN to evolving SIR properties. Panels (a-e) demonstrate the continuous ingestion of solar wind data. The model probability (black) and catalogue SIR interval (red shaded region) are shown along with velocity (blue) and density (green) time series.
    \label{fig:edge_performance}}
    \vspace{-0.5cm}
\end{figure*}

The analyses in Sections 4.1 and 4.2 evaluate SIREN on complete 6-day sequences with the SIR centered in the observation window, flanked by ambient slow wind upstream and fast wind downstream. During training, the model is exposed to this configuration, with fully resolved SIR bracketed by pre- and post-event context. However, in an operational setting, solar wind data arrives continuously, and the SIR develops gradually. Therefore, it is essential to verify that SIREN produces meaningful probability estimates when the SIR is not centered and the full event context has not accumulated.

Figure \ref{fig:edge_performance} presents this validation by simulating real-time ingestion of solar wind data for a representative SIR event from test set. Panels (a) through (e) show successive snapshots separated by approximately 9 hours, in which progressively more data is available to the model. Although only velocity and density are displayed for visual clarity, the model ingests all 11 input features (including magnetic field components, temperature, and flow angles) at every timestep. The model probability reaches its peak values within the catalog-defined SIR interval and subsequently declines as the post-interface fast wind dominates, producing a well-defined probability profile similar to those obtained from the centered test-set evaluation.

The model's probability output increases monotonically as the SIR develops within the observation window: from near-zero (panel a) during the ambient phase, through intermediate values as early compression signatures appear (panel b and c), to high confidence once the bulk of the interaction region is observed (panel d and e). This graduated response is desirable for real-time monitoring, as it provides an early indication of an emerging SIR before the full event has passed the spacecraft.

This progressive behavior is noteworthy because the model was never explicitly trained on partial or edge-positioned SIR sequences. The results in Figure \ref{fig:edge_performance} demonstrate that SIREN responds to the physical content of the input rather than to its absolute position within the window. This suggests that the self-attention mechanism, which computes pairwise relationships between all available timesteps, enables the model to recognize compression and deflection signatures regardless of where they fall in the sequence. The positional embedding provides temporal ordering information that helps the model parse the sequential structure of the wind, but it does not rigidly bind detection to a fixed position.

These findings support the potential deployment of SIREN in operational space weather pipelines, where real-time inference on continuously updating data streams is the primary use case.

\section{Discussion}

\subsection{Manual Detection vs. SIREN}
Traditional SIR catalogs rely on expert inspection of time series of in-situ data, where an analyst identifies the onset and offset of stream interaction based on visual assessment of velocity gradients, density enhancements, magnetic field compression, and other variations. This process is inherently subjective: two experts examining the same event may place the SIR boundaries at different timesteps, particularly at the leading and trailing edges where the transition between ambient wind and compressed plasma is gradual. The resulting catalogs are binary by construction and offer no measure of confidence in the boundary placement.

The Transformer-based approach presented here addresses both limitations. Rather than forcing a binary decision at every timestep, the model produces a continuous probability of SIR membership. Events with clear compression signatures receive probabilities near unity, while ambiguous boundary regions naturally receive intermediate values. This probabilistic framing is operationally valuable: a space weather forecaster can apply a conservative threshold (e.g., 0.8) when false alarms are costly, or a permissive threshold (e.g., 0.4) when comprehensive event coverage is the priority. The choice of threshold becomes an explicit, adjustable decision rather than an implicit consequence of expert judgment.

\begin{figure}[h]
    \includegraphics[width = \textwidth]{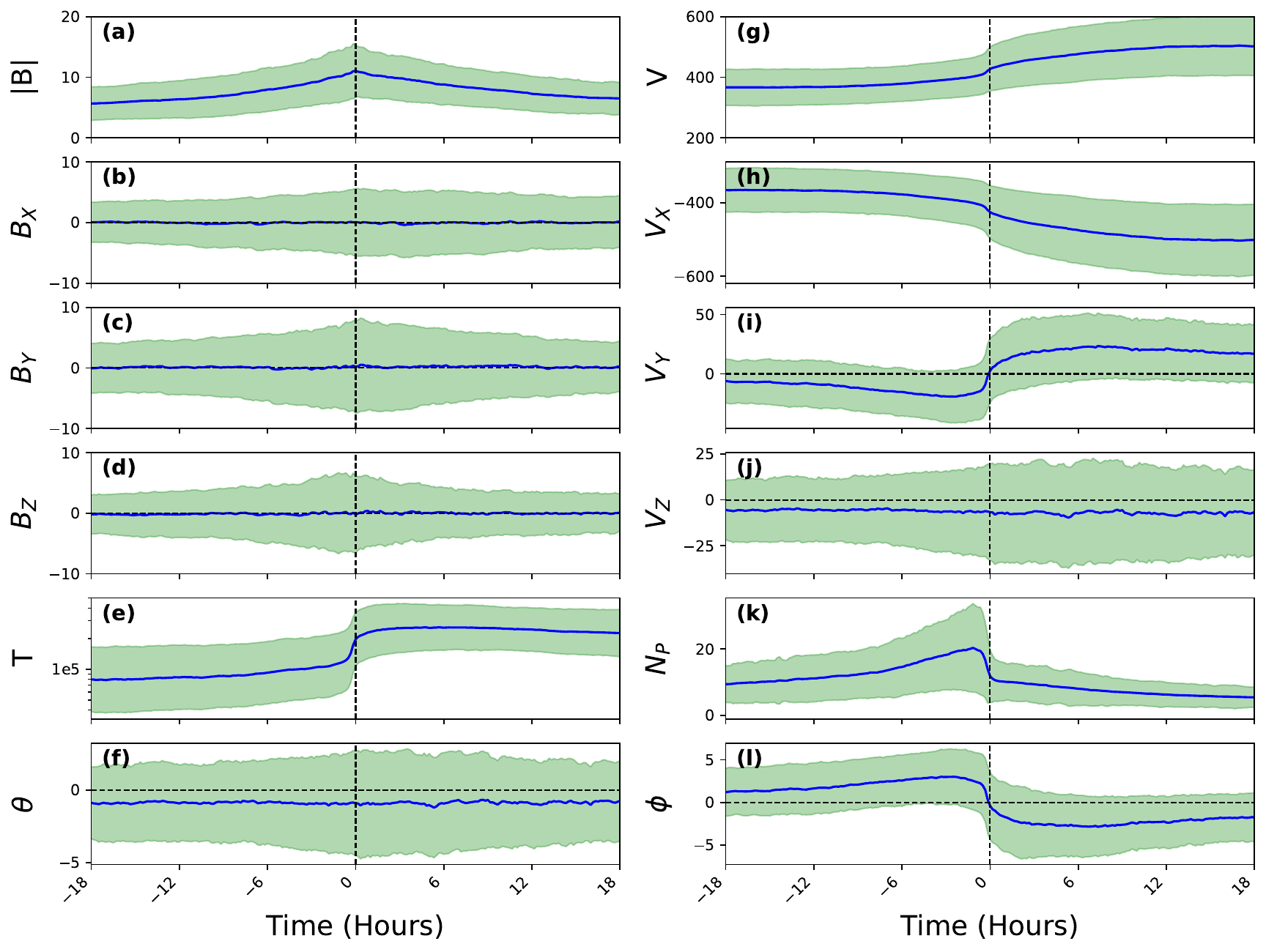}
    \caption{Superposed epoch analysis (SPA) of 676 SIR events. The black vertical dashed line denotes the epoch time, defined by the stream interface. Panels (a–f) in the left column show the magnetic field magnitude and components ($|B|$, $B_x$, $B_y$, $B_z$), temperature ($T$), and the north–south flow angle ($\theta$). Panels (g–l) in the right column display the velocity magnitude and components ($|V|$, $V_x$, $V_y$, $V_z$), proton number density ($N_P$), and the east–west bulk flow angle ($\phi$). The blue line represents the mean, and the green shaded region indicates the 1-$\sigma$ variation.
    \label{fig:spa}}
    \vspace{-1.5cm}
\end{figure}

Beyond the probabilistic framework, the model's per-timestep probability profiles reveal two behaviors that go beyond what binary catalogs can capture. First, in several test-set events the model probability begins rising 10-30 timesteps (roughly 2-5 hours) before the catalog-defined SIR onset. In these cases, the corresponding solar wind data show early density perturbations or initial velocity increases that precede the catalog boundary. The fact that the model was trained on the same catalog labels yet learned to anticipate onset earlier suggests it has generalized beyond the specific boundaries it was shown, responding instead to the underlying physical signatures of incipient compression.

Second, the model exhibits mid-event probability modulations within several catalog-defined SIR windows, where the probability temporarily decreases before recovering. These dips correspond to local decreases in density or magnetic field strength within the interaction region, intervals where the plasma conditions temporarily depart from the compression signature. A manual catalog treats the entire SIR window as uniformly labeled, but the model recognizes that the local plasma state has momentarily relaxed. This behavior demonstrates that the model continuously evaluates the input features rather than applying a simple on/off classification once the SIR onset is detected, providing a more nuanced characterization of stream interaction than binary labeling can offer.

\subsection{Superposed Epoch Validation}

Figure \ref{fig:spa} presents the superposed epoch analysis (SPA) of 676 SIR events, where the stream interface is used as the reference epoch (indicated by the black vertical dashed line at t = 0). In all panels, the blue curve represents the mean profile obtained from the ensemble of events, while the green shaded region denotes the 1-sigma variability, providing a measure of event-to-event fluctuations.

The magnetic field magnitude ($|B|$) exhibits a clear enhancement near the stream interface, consistent with the compression region formed by the interaction of fast and slow solar wind streams. The proton temperature ($T$) increases around the stream interface, indicating plasma heating within the compression region. The magnetic field components ($B_x$, $B_y$, $B_z$) and the north--south flow angle ($\theta$) display relatively steady mean values across the interaction region.

The velocity magnitude ($|V|$) shows a gradual increase across the stream interface, marking the transition from slow to fast solar wind. The velocity components ($V_x$ and $V_y$) reveal directional changes in the flow, highlighting the complex dynamics of the interaction region, whereas $V_z$ remains nearly constant. The proton number density ($N_P$) peaks near the stream interface, consistent with plasma compression. The east--west GSE flow angle ($\phi$) reverses across the interface, indicating longitudinal deflections in the solar wind flow.

A natural concern with any machine learning model is whether it has learned genuine physical patterns or merely memorized artifacts of its training data. To address this, we compare the model's learned feature hierarchy against the superposed epoch analysis of the full SIR dataset, which reveals the average physical signatures of stream interaction regions independent of the model.
The superposed median profiles show several well-established features of SIRs at 1 AU. Total magnetic field magnitude $|B|$ peaks at the stream interface. Proton density (N$_p$) peaks just ahead of the interface before declining into the less compressed fast wind. Bulk speed ($V$) transitions from 350–400 km/s to 500+ km/s across the interface. Temperature rises in the post-interface high-speed stream. These are the classical signatures of stream interaction, and they correspond directly to the top-ranked features in our attribution analysis.

What is more notable is the behavior of the secondary features. The superposed profiles of the east-west flow angle ($\phi$) and transverse velocity (V$_y$) both show clear, systematic deflections centered on the stream interface. The flow angle exhibits a negative excursion of 2–3 degrees, consistent with the westward deflection of solar wind plasma compressed against the stream interface. V$_y$ shows a corresponding perturbation of 10–20 km/s. These signatures are present in the superposed statistics, yet they are rarely emphasized in manual catalog definitions, which focus primarily on density and magnetic field compression.

SIREN is trained without any explicit instruction about flow deflection and independently assigned 7-9\% importance to the east-west flow angle and 6-8\% to $V_y$. This convergence between the data-driven feature ranking and the superposed physical signatures provides strong evidence that the transformer learned genuine SIR properties. This observation illustrates a broader point about the relationship between manual catalogs and statistical learning. A human expert examining individual events may notice flow deflection in some cases but not others, and may not weight it consistently across events because it is subtle compared to the dramatic density and field enhancements. SIREN, by contrast, processes all events uniformly and aggregates weak but consistent signals across the full training set.

\subsection{Physics Insight}
The consistent importance of V$_y$ and the east-west flow angle in the model's feature hierarchy carries implications beyond SIR detection. These transverse flow properties arise from the three-dimensional interaction between fast and slow solar wind streams. As the compression region develops, plasma is deflected both radially and tangentially, producing systematic perturbations in the flow angle and transverse velocity components. This process is well captured by 3D MHD simulations, which naturally reproduce flow deflection at stream interfaces as a consequence of the longitudinal pressure gradient across the compression region. Prior simulation study \cite{mayank_2022_swastisw} and observational analysis \cite{Gosling1978, Gosling1999, Rout2017, Yogesh2023} reported flow deflection at SIRs, but these findings had not been quantitatively ranked against other SIR properties in a systematic manner. The present work provides that ranking: V$_y$ and the east-west flow angle together account for approximately 13–17\% of the total feature importance, placing them firmly above noise level and below the dominant compression indicators.

This finding also highlights a distinction between solar wind modeling approaches. Simple extrapolation schemes, such as Heliospheric Upwinding eXtrapolation (HUX) \cite{riley_2011_hux} and similar radial propagation models, can predict bulk speed variations and, to some extent, density compression through kinematic steepening, but they cannot produce transverse velocity deflections or flow angle perturbations because they do not resolve three-dimensional pressure gradients. The systematic importance of V$_y$ and the flow angle in our attribution analysis suggests that these simplified models miss a non-trivial component of the SIR signature. MHD models \cite{2014ApJ...782...81V, sunrunner3D, Mayank_2025_swasti}, by contrast, naturally produce these transverse features as part of the self-consistent solution.

More broadly, the per-sample attribution analysis reveals that the model adapts its feature weighting depending on event morphology. In strong SIRs, dramatic density and field compression can dominate with 40-45\% importance. In weaker events where compression is modest, V$_y$ and the flow angles gain relative importance, rising to 8–10\% each. This adaptive behavior suggests that SIRs are not monolithic structures with a fixed signature hierarchy; rather, the relative diagnostic value of different properties varies with event strength and geometry. This observation opens a path toward event subtyping based on feature importance profiles, which could inform both forecasting strategies and our physical understanding of stream interaction variability.

\subsection{Limitations and Follow Ups}
There are a few limitations of the current study which should be acknowledged. First, the model is trained exclusively on SIR events identified in solar wind data free of significant CME contamination. In practice, coronal mass ejections frequently overlap with stream interaction regions, producing complex plasma signatures that blend CME and SIR characteristics. The current model has not been exposed to such mixed events during training and is likely to produce unreliable predictions when CME-associated structures are present in the input sequence. Extending the framework to handle CME-contaminated intervals, or to jointly classify SIRs and CMEs within the same sequence, is a natural next step.

Second, the model is trained and evaluated on \textit{Wind} data at 1 AU. The physical properties of solar wind \cite{Yogesh2026} as well as SIRs evolve with heliocentric distance: compression strengthens, shocks may form, and the relative importance of different features likely shifts. Validating the learned feature hierarchy against observations from STEREO, Parker Solar Probe, Solar Orbiter, and MAVEN at different radial distances would test whether the rankings reported here are intrinsic to stream interaction physics or specific to the 1 AU vantage point.

Finally, the interpretability framework demonstrated here for SIRs is not limited to stream interaction regions. The same pipeline of training a compact neural network, applying gradient-based attribution, and validating against superposed physical observations can be applied to other space weather event types. CME sheath detection, magnetic cloud identification, and solar wind type classification (slow, fast, CIR, transient) are all amenable to this approach. In each case, the method would reveal which in-situ properties the model relies on for detection, providing quantitative insight into the feature hierarchies that define these structures. We propose that feature attribution analysis should become standard practice in space weather machine learning, complementing traditional accuracy metrics with physical interpretability.

\section{Conclusion}

This work presents the first application of a lightweight transformer architecture to stream interaction region detection in in-situ solar wind observations, combined with systematic feature attribution analysis to quantify which physical properties define SIRs.
The model achieves robust detection performance across diverse event morphologies, with a receiver operating characteristic area under the curve of 0.93, an F1 score of 0.78, and a true skill statistic of 0.67 on an independent test set of 102 events. The transformer's self-attention mechanism concentrates on the stream interface region, confirming that the model identifies the physically relevant portion of each sequence.

Feature attribution reveals a quantifiable hierarchy of SIR properties. Proton density and total magnetic field magnitude together account for approximately 46–50\% of the total attribution, consistent with their role as the primary compression indicators. Bulk speed and temperature contribute 22–26\%, reflecting the velocity transition and post-interface heating. Notably, the transverse velocity component V$_y$ and the east-west flow angle together account for 13-17\%, confirming their role as consistent secondary SIR signatures. This quantitative validation of flow deflection as a diagnostic feature, previously reported in MHD simulations and select observational studies, had not been established through systematic data-driven analysis.

Per-sample attribution analysis demonstrates that the model adapts its feature weighting by event morphology: strong compression events are dominated by density and magnetic field, while weaker events rely proportionally more on flow structure. This adaptive behavior reflects genuine physical variability in stream interaction signatures.
These contributions carry four broader implications for the field:

\textit{Probabilistic catalog formation}: Continuous probability outputs replace binary decisions, enabling automated SIR catalog construction where the detection threshold can be used to classify different types of SIRs.

\textit{Operational deployment}: With $<$0.1 million parameters, the model is compact enough for real-time inference in operational space weather pipelines and for potential onboard deployment on future heliospheric missions.

\textit{Transferable interpretability}: The same train-attribute-validate pipeline is directly applicable to other space weather classification tasks, offering a reusable template for physics-grounded interpretability in heliophysics.

\textit{Cross-mission validation}: Testing the learned feature hierarchy against observations from multiple spacecraft at different heliocentric distances would establish whether the rankings reported here are universal properties of stream interaction physics.

Looking ahead, extending SIREN to handle CME-contaminated intervals, distinguishing between corotating and transient interaction regions, and applying the same interpretability pipeline to multi-mission datasets represent natural next steps toward establishing data-driven SIR characterization as a robust, mission-independent capability.

%
%
%
%

%



%
%

\section*{Open Research Section}
To support reproducibility and community uptake, all source code, trained model checkpoints (including weights and biases), and the feature attribution pipeline are made publicly available at https://doi.org/10.5281/zenodo.20531966.

\section*{Conflict of Interest declaration}
The authors declare there are no conflicts of interest for this manuscript.

\acknowledgments
This research was supported by the NASA Living with a Star Jack Eddy Postdoctoral Fellowship Program, administered by UCAR’s Cooperative Programs for the Advancement of Earth System Science (CPAESS) under award No. 80NSSC22M0097.  This work was partially supported by NASA under awards No. 80NSSC23M0192, 80NSSC20K1580, 80NSSC21K1555.
Y. and L.K.J. acknowledge support by NASA Heliophysics Guest Investigator Grant 80NSSC23K0447.
Y. acknowledges the support by the College of Liberal Arts and Sciences at the University of Iowa. G.G.H. was supported by NASA grant 80NSSC24K0552.
We gratefully acknowledge the principal investigators of instruments on-board NASA's Wind mission for generating the data and making them publicly available. We acknowledge use of NASA/GSFC’s Space Physics Data Facility (SPDF).

\appendix
\section{Feature Attribution via Integrated Gradients}
To interpret which input features drive the model's SIR detection decisions, we apply Integrated Gradients (IG) \cite{Sundararajan_2017}, a gradient-based attribution method with two key properties. First, completeness: the sum of IG attributions across all input features exactly equals the difference in model output between the actual input and a reference baseline. Second, sensitivity: if a feature differs between the input and the baseline, and this difference changes the model output, the feature receives nonzero attribution. These axiomatic guarantees make IG preferable to simpler gradient methods (e.g., vanilla saliency maps) that lack completeness, and to perturbation-based methods like SHAP (SHapley Additive exPlanations), that are prohibitively expensive for high-dimensional sequential inputs.

IG computes the attribution for each input feature by integrating the model's gradient along a straight-line path from a reference baseline x' to the actual input x:

\begin{equation}
    IG_i(x) = (x_i - x'_i) \times \int (\delta F/\delta x_i)(x' + \alpha(x - x')) \quad d\alpha
\end{equation}

where F is the model output, i indexes the input feature, and $\alpha$ parameterizes the interpolation from baseline to input. In practice, the integral is approximated by a Riemann sum over a discrete number of interpolation steps. We use 128 steps, which yields mean convergence deltas below 0.01 across the test set, indicating accurate numerical integration.

\subsection{Baseline and Target}
The choice of reference baseline is critical for IG interpretation. The baseline should represent an ``uninformative" input state from which the model would produce a low (near-zero) SIR probability. We use a pre-event baseline defined as the mean of the first 120 timesteps of each sequence (approximately the first 20 hours). Because each sequence is constructed with the SIR event positioned toward the center, these leading timesteps correspond to ambient slow solar wind preceding the compression onset. This baseline is physically motivated: it represents the quiet-wind conditions from which SIR signatures emerge, and it is well-defined for all normalization schemes.
For the IG target, we wrap the model to produce a scalar output by summing the raw logits (pre-sigmoid) over the SIR-labeled timesteps only. Restricting the summation to the SIR window avoids diluting the attribution signal with quiet-wind timesteps where both the input and the baseline produce near-zero gradients. If no SIR label is present for a given sample, the summation falls back to the full sequence. Using raw logits rather than sigmoid probabilities is standard practice in the attribution literature, as sigmoid saturation in high-confidence regions compresses gradients and obscures feature contributions.

\subsection{Aggregation and Interpretation}
We compute IG attributions for all 102 test-set samples. For each sample, the attribution tensor has shape [864, 11], assigning a signed value to every feature at every timestep. We aggregate these attributions in three ways.

First, global feature importance: for each sample, we compute the mean absolute attribution per feature (averaged over all 864 timesteps), then average across samples. Absolute values ensure that features with large positive and large negative attributions are recognized as important, regardless of sign. The resulting per-feature importance scores are sum-normalized to fractional contributions (percentages). We also compute a SIR-window-only variant by restricting the temporal average to labeled SIR timesteps.

Second, temporal attribution maps: we average the attribution magnitude (and separately, the signed attribution) at each timestep across all test samples, producing heatmaps that reveal where in the sequence each feature contributes to detection. These maps are normalized to [0, 1] for magnitude and [-1, 1] for signed direction.

Third, per-sample case studies: for selected events, we visualize the raw solar wind data, model probability output, signed attribution heatmap, and feature importance bar chart together, enabling direct comparison of physical signatures with learned attributions.
The sign of IG attribution carries physical meaning. Positive attribution at a given timestep indicates that the feature value at that timestep, relative to the baseline, pushes the model output toward higher SIR probability. Negative attribution indicates the feature pushes the model away from SIR detection. A feature can be highly important (large absolute attribution) while exhibiting mixed or net-negative signed attribution if its temporal dynamics within the SIR involve both enhancements and local decreases relative to the baseline. We discuss this in detail for proton density in Section 4.

%
\bibliography{Sir} 

@ARTICLE{Lepping1995,
       author = {{Lepping}, R.~P. and {Ac{\~{u}}na}, M.~H. and {Burlaga}, L.~F. and {Farrell}, W.~M. and {Slavin}, J.~A. and {Schatten}, K.~H. and {Mariani}, F. and {Ness}, N.~F. and {Neubauer}, F.~M. and {Whang}, Y.~C. and {Byrnes}, J.~B. and {Kennon}, R.~S. and {Panetta}, P.~V. and {Scheifele}, J. and {Worley}, E.~M.},
        title = "{The Wind Magnetic Field Investigation}",
      journal = {Space Science Reviews},
         year = 1995,
        month = feb,
       volume = {71},
       number = {1-4},
        pages = {207-229},
          doi = {10.1007/BF00751330},
       adsurl = {https://ui.adsabs.harvard.edu/abs/1995SSRv...71..207L}
}

@ARTICLE{Ogilvie1995,
       author = {{Ogilvie}, K.W. and {Chornay}, D.J. and {Fritzenreiter}, R.J. and {Hunsaker}, F. and {Keller}, J. and {Lobell}, J. and {Miller}, G. and {Scudder}, JD. and {Sittler}, Jr., E.C. and {Torbert}, R.B. and {Bodet}, D. and {Needell}, G. and {Lazarus}, A.J. and {Steinberg}, J.T. and {Tappan}, J.H. and {Mavretic}, A. and {Gergin}, E.},
        title = "{SWE, A Comprehensive Plasma Instrument for the Wind Spacecraft}",
      journal = {Space Science Reviews},
         year = 1995,
        month = feb,
       volume = {71},
       number = {1-4},
        pages = {55-77},
          doi = {10.1007/BF00751326},
       adsurl = {https://ui.adsabs.harvard.edu/abs/1995SSRv...71...55O}
}

@ARTICLE{Chi2018,
       author = {{Chi}, Yutian and {Shen}, Chenglong and {Luo}, Bingxian and {Wang}, Yuming and {Xu}, Mengjiao},
        title = "{Geoeffectiveness of Stream Interaction Regions From 1995 to 2016}",
      journal = {Space Weather},
         year = 2018,
        month = dec,
       volume = {16},
       number = {12},
        pages = {1960-1971},
          doi = {10.1029/2018SW001894},
       adsurl = {https://ui.adsabs.harvard.edu/abs/2018SpWea..16.1960C}
}

@ARTICLE{Yogesh2023,
       author = {{Yogesh} and {Chakrabarty}, D. and {Srivastava}, Nandita},
        title = "{New insights on the behaviour of solar wind protons and alphas in the stream interaction region in solar cycle 23 and 24}",
      journal = {Monthly Notices of the Royal Astronomical Society},
         year = 2023,
        month = nov,
       volume = {526},
       number = {1},
        pages = {L13-L19},
          doi = {10.1093/mnrasl/slad112},
archivePrefix = {arXiv},
       eprint = {2304.00274},
 primaryClass = {astro-ph.SR},
       adsurl = {https://ui.adsabs.harvard.edu/abs/2023MNRAS.526L..13Y}
}

@ARTICLE{Alves2006,
       author = {{Alves}, M.~V. and {Echer}, E. and {Gonzalez}, W.~D.},
        title = "{Geoeffectiveness of corotating interaction regions as measured by Dst index}",
      journal = {Journal of Geophysical Research (Space Physics)},
         year = 2006,
        month = jul,
       volume = {111},
       number = {A7},
          eid = {A07S05},
        pages = {A07S05},
          doi = {10.1029/2005JA011379},
       adsurl = {https://ui.adsabs.harvard.edu/abs/2006JGRA..111.7S05A}
}

@ARTICLE{Balogh2013,
       author = {{Balogh}, Andr{\'e} and {Erd{\~o}s}, G{\'e}za},
        title = "{The Heliospheric Magnetic Field}",
      journal = {Space Science Reviews},
         year = 2013,
        month = jun,
       volume = {176},
       number = {1-4},
        pages = {177-215},
          doi = {10.1007/s11214-011-9835-3},
       adsurl = {https://ui.adsabs.harvard.edu/abs/2013SSRv..176..177B}
}

@ARTICLE{Jian2006,
       author = {{Jian}, L. and {Russell}, C.~T. and {Luhmann}, J.~G. and {Skoug}, R.~M.},
        title = "{Properties of Stream Interactions at One AU During 1995   2004}",
      journal = {Solar Physics},
         year = 2006,
        month = dec,
       volume = {239},
       number = {1-2},
        pages = {337-392},
          doi = {10.1007/s11207-006-0132-3},
       adsurl = {https://ui.adsabs.harvard.edu/abs/2006SoPh..239..337J}
}

@ARTICLE{Echer2013,
       author = {{Echer}, E. and {Tsurutani}, B.~T. and {Gonzalez}, W.~D.},
        title = "{Interplanetary origins of moderate (-100 nT < Dst {\ensuremath{\leq}} -50 nT) geomagnetic storms during solar cycle 23 (1996-2008)}",
      journal = {Journal of Geophysical Research (Space Physics)},
         year = 2013,
        month = jan,
       volume = {118},
       number = {1},
        pages = {385-392},
          doi = {10.1029/2012JA018086},
       adsurl = {https://ui.adsabs.harvard.edu/abs/2013JGRA..118..385E}
}

@ARTICLE{Tsurutani2006,
       author = {{Tsurutani}, Bruce T. and {Gonzalez}, Walter D. and {Gonzalez}, Alicia L.~C. and {Guarnieri}, Fernando L. and {Gopalswamy}, Nat and {Grande}, Manuel and {Kamide}, Yohsuke and {Kasahara}, Yoshiya and {Lu}, Gang and {Mann}, Ian and {McPherron}, Robert and {Soraas}, Finn and {Vasyliunas}, Vytenis},
        title = "{Corotating solar wind streams and recurrent geomagnetic activity: A review}",
      journal = {Journal of Geophysical Research (Space Physics)},
         year = 2006,
        month = jul,
       volume = {111},
       number = {A7},
          eid = {A07S01},
        pages = {A07S01},
          doi = {10.1029/2005JA011273},
       adsurl = {https://ui.adsabs.harvard.edu/abs/2006JGRA..111.7S01T}
}

@ARTICLE{Gosling1999,
       author = {{Gosling}, J.~T. and {Pizzo}, V.~J.},
        title = "{Formation and Evolution of Corotating Interaction Regions and their Three Dimensional Structure}",
      journal = {Space Science Reviews},
         year = 1999,
        month = jul,
       volume = {89},
        pages = {21-52},
          doi = {10.1023/A:1005291711900},
       adsurl = {https://ui.adsabs.harvard.edu/abs/1999SSRv...89...21G}
}

@ARTICLE{Zhang2008,
       author = {{Zhang}, Y. and {Sun}, W. and {Feng}, X.~S. and {Deehr}, C.~S. and {Fry}, C.~D. and {Dryer}, M.},
        title = "{Statistical analysis of corotating interaction regions and their geoeffectiveness during solar cycle 23}",
      journal = {Journal of Geophysical Research (Space Physics)},
         year = 2008,
        month = aug,
       volume = {113},
       number = {A8},
          eid = {A08106},
        pages = {A08106},
          doi = {10.1029/2008JA013095},
       adsurl = {https://ui.adsabs.harvard.edu/abs/2008JGRA..113.8106Z}
}

@ARTICLE{Lockwood2016,
       author = {{Lockwood}, Mike and {Owens}, Mathew J. and {Barnard}, Luke A. and {Bentley}, Sarah and {Scott}, Chris J. and {Watt}, Clare E.},
        title = "{On the origins and timescales of geoeffective IMF}",
      journal = {Space Weather},
         year = 2016,
        month = jun,
       volume = {14},
       number = {6},
        pages = {406-432},
          doi = {10.1002/2016SW001375},
       adsurl = {https://ui.adsabs.harvard.edu/abs/2016SpWea..14..406L}
}

@ARTICLE{Jian2011,
       author = {{Jian}, L.~K. and {Russell}, C.~T. and {Luhmann}, J.~G.},
        title = "{Comparing Solar Minimum 23/24 with Historical Solar Wind Records at 1 AU}",
      journal = {Solar Physics},
         year = 2011,
        month = dec,
       volume = {274},
       number = {1-2},
        pages = {321-344},
          doi = {10.1007/s11207-011-9737-2},
       adsurl = {https://ui.adsabs.harvard.edu/abs/2011SoPh..274..321J}
}

@ARTICLE{Yogesh2021,
       author = {{Yogesh} and {Chakrabarty}, D. and {Srivastava}, N.},
        title = "{Evidence for distinctive changes in the solar wind helium abundance in solar cycle 24}",
      journal = {Monthly Notices of the Royal Astronomical Society},
         year = 2021,
        month = may,
       volume = {503},
       number = {1},
        pages = {L17-L22},
          doi = {10.1093/mnrasl/slab016},
archivePrefix = {arXiv},
       eprint = {2102.05395},
 primaryClass = {astro-ph.SR},
       adsurl = {https://ui.adsabs.harvard.edu/abs/2021MNRAS.503L..17Y}
}

@ARTICLE{Yogesh2024,
       author = {{Yogesh} and {Gopalswamy}, N. and {Chakrabarty}, D. and {Mostafavi}, Parisa and {Yashiro}, Seiji and {Srivastava}, Nandita and {Ofman}, Leon},
        title = "{Origins of Very Low Helium Abundance Streams Detected in the Solar Wind Plasma}",
      journal = {The Astrophysical Journal},
         year = 2024,
        month = dec,
       volume = {977},
       number = {1},
          eid = {89},
        pages = {89},
          doi = {10.3847/1538-4357/ad84d6},
archivePrefix = {arXiv},
       eprint = {2410.04713},
 primaryClass = {astro-ph.SR},
       adsurl = {https://ui.adsabs.harvard.edu/abs/2024ApJ...977...89Y}
}

@article{Ofman2024,
doi = {10.3847/2041-8213/ad5e7e},
url = {https://dx.doi.org/10.3847/2041-8213/ad5e7e},
year = {2024},
month = {jul},
publisher = {The American Astronomical Society},
volume = {970},
number = {1},
pages = {L16},
author = {Leon Ofman and  Yogesh and Silvio Giordano},
title = {Understanding the Variability of Helium Abundance in the Solar Corona Using Three-fluid Modeling and Ultraviolet Observations},
journal = {The Astrophysical Journal Letters},
}

@article{Durovcova2019,
  title={Evolution of relative drifts in the expanding solar wind: Helios observations},
  author={{\v{D}}urovcov{\'a}, Tereza and {\v{S}}afr{\'a}nkov{\'a}, Jana and N{\v{e}}me{\v{c}}ek, Zden{\v{e}}k},
  journal={Solar Physics},
  volume={294},
  number={7},
  pages={97},
  year={2019},
  publisher={Springer}
}

@ARTICLE{Khaled_enrico_2023,
       author = {{Alielden}, Khaled and {Camporeale}, Enrico and {Kors{\'o}s}, Marianna B. and {Taroyan}, Youra},
        title = "{Prediction Interval of Interface Regions: Machine Learning Nowcasting Approach}",
      journal = {Space Weather},
         year = 2023,
        month = mar,
       volume = {21},
       number = {3},
          eid = {e2022SW003326},
        pages = {e2022SW003326},
          doi = {10.1029/2022SW003326},
       adsurl = {https://ui.adsabs.harvard.edu/abs/2023SpWea..2103326A}
}

@article{mayank_2022_swastisw,
  author = {Mayank, Prateek and Vaidya, Bhargav and Chakrabarty, D.},
  month = {09},
  pages = {23},
  title = {SWASTi-SW: Space Weather Adaptive Simulation Framework for Solar Wind and Its Relevance to the Aditya-L1 Mission},
  doi = {10.3847/1538-4365/ac8551},
  url = {https://doi.org/10.3847/1538-4365/ac8551},
  urldate = {2023-01-31},
  volume = {262},
  year = {2022},
  journal = {The Astrophysical Journal Supplement Series}
}

@article{prateekmayank_2023_swasticme,
  author = {Prateek Mayank and Vaidya, Bhargav and Mishra, Wageesh and Chakrabarty, D},
  month = {12},
  pages = {10-10},
  publisher = {Institute of Physics},
  title = {SWASTi-CME: A Physics-based Model to Study Coronal Mass Ejection Evolution and Its Interaction with Solar Wind},
  doi = {10.3847/1538-4365/ad08c7},
  urldate = {2024-05-03},
  volume = {270},
  year = {2023},
  journal = {The Astrophysical Journal Supplement Series}
}

@article{Mayank2024,
       author = {{Mayank}, Prateek and {Lotz}, Stefan and {Vaidya}, Bhargav and {Mishra}, Wageesh and {Chakrabarty}, D.},
        title = "{Study of Evolution and Geo-effectiveness of Coronal Mass Ejection{\textendash}Coronal Mass Ejection Interactions Using Magnetohydrodynamic Simulations with SWASTi Framework}",
      journal = {The Astrophysical Journal},
         year = 2024,
        month = nov,
       volume = {976},
       number = {1},
          eid = {126},
        pages = {126},
          doi = {10.3847/1538-4357/ad8084},
archivePrefix = {arXiv},
       eprint = {2409.19943},
 primaryClass = {astro-ph.SR},
       adsurl = {https://ui.adsabs.harvard.edu/abs/2024ApJ...976..126M}
}

@article{Smitha2025,
       author = {{Thampi}, Smitha V. and {Bhaskar}, Ankush and {Mayank}, Prateek and {Vaidya}, Bhargav and {Venugopal}, Indu},
        title = "{Simulating the Arrival of Multiple Coronal Mass Ejections That Triggered the Gannon Superstorm on 2024 May 10}",
      journal = {The Astrophysical Journal},
         year = 2025,
        month = mar,
       volume = {981},
       number = {1},
          eid = {76},
        pages = {76},
          doi = {10.3847/1538-4357/ada93c},
archivePrefix = {arXiv},
       eprint = {2411.08612},
 primaryClass = {physics.space-ph},
       adsurl = {https://ui.adsabs.harvard.edu/abs/2025ApJ...981...76T}
}

@article{kay2024,
  title={Updating measures of CME arrival time errors},
  author={Kay, C and Palmerio, E and Riley, P and Mays, ML and Nieves-Chinchilla, T and Romano, M and Collado-Vega, YM and Wiegand, C and Chulaki, A},
  journal={Space Weather},
  volume={22},
  number={7},
  pages={e2024SW003951},
  year={2024},
  publisher={Wiley Online Library}
}

@ARTICLE{2026esoar.92354528C,
       author = {{Corti}, Claudio and {Kuznetsova}, Maria M. and {Reiss}, Martin and {Yue}, Jia and {Karpen}, Judith T. and {Arge}, Charles Nickolos and {Bacchini}, Fabio and {Bard}, Christopher and {Bruinsma}, Sean L. and {Caplan}, Ronald M. and et al.},
        title = "{Advancing Heliophysics and Space Weather Modeling through Open Science}",
      journal = {ESS Open Archive eprints},
         year = 2026,
        month = jan,
       volume = {923},
        pages = {54528},
          doi = {10.22541/essoar.176824639.92354528/v1},
       adsurl = {https://ui.adsabs.harvard.edu/abs/2026esoar.92354528C}
}

@ARTICLE{Hajra_2022,
       author = {{Hajra}, Rajkumar and {Sunny}, Jibin V.},
        title = "{Corotating Interaction Regions during Solar Cycle 24: A Study on Characteristics and Geoeffectiveness}",
      journal = {Solar Physics},
         year = 2022,
        month = mar,
       volume = {297},
       number = {3},
          eid = {30},
        pages = {30},
          doi = {10.1007/s11207-022-01962-1},
       adsurl = {https://ui.adsabs.harvard.edu/abs/2022SoPh..297...30H}
}

@ARTICLE{Vishal_2020,
       author = {{Upendran}, Vishal and {Cheung}, Mark C.~M. and {Hanasoge}, Shravan and {Krishnamurthi}, Ganapathy},
        title = "{Solar Wind Prediction Using Deep Learning}",
      journal = {Space Weather},
         year = 2020,
        month = sep,
       volume = {18},
       number = {9},
          eid = {e02478},
        pages = {e02478},
          doi = {10.1029/2020SW002478},
archivePrefix = {arXiv},
       eprint = {2006.05825},
 primaryClass = {astro-ph.SR},
       adsurl = {https://ui.adsabs.harvard.edu/abs/2020SpWea..1802478U}
}

@ARTICLE{Indian_helio2025,
       author = {{Nandy}, Dibyendu and {Pant}, Vaibhav and {Anand}, Megha and {Athalathil}, Jithu J. and {Awasthi}, Arun Kumar and {Bane}, Kshitij and {Banerjee}, Dipankar and {Ravindra}, B. and {Bhaskar}, Ankush and {Bhattacharyya}, R. and {Bhowmik}, Prantika and {Chandra}, Ramesh and {Chatterjee}, Piyali and {Chatterjee}, Subhamoy and {Dimri}, A.~P. and {Gokani}, Sneha A. and {Hanasoge}, Shravan and {Hazra}, Soumitra and {Jain}, Rajmal and {Joshi}, Bhuvan and {Nagaraju}, K. and {Kansabanik}, Devojyoti and {Karak}, Bidya Binay and {Kathiravan}, C. and {Khan}, Raveena and {Krishnan}, Hariharan and {Kumar}, Brajesh and {Kumar}, Sanjay and {Kumari}, Anshu and {Majumdar}, Satabdwa and {Mayank}, Prateek and {Mishra}, Sudheer and {Mishra}, Wageesh and {Mohan}, Atul and {Mondal}, Surajit and {Mugundhan}, V. and {Narendranath}, Shyama and {Oberoi}, Divya and {Pandya}, Megha and {Patel}, Ritesh and {Paul}, Arghyadeep and {Prasad}, Avijeet and {Raja}, K. Sasikumar and {Rajhans}, Abhishek and {Ramesh}, R. and {Saha}, Chitradeep and {Sankarasanubramanian}, K. and {Selvakumaran}, R. and {Sharma}, Rahul and {Sharma}, Rohit and {Shrivastav}, Arpit Kumar and {Singh}, Nishant and {Lata Soni}, Shirsh and {Srivastava}, Abhishek K. and {Srivastava}, Nandita and {Tripathi}, Durgesh and {Uddin}, Wahab and {Vaidya}, Bhargav and {Vemareddy}, P. and {Vichare}, Geeta and {Vigeesh}, Gangadharan and {Yadav}, Nitin and {Yadav}, Vipin K.},
        title = "{Indian solar and heliospheric physics vision: Fundamental science to a space weather resilient society}",
      journal = {Journal of Astrophysics and Astronomy},
         year = 2025,
        month = aug,
       volume = {46},
       number = {2},
          eid = {51},
        pages = {51},
          doi = {10.1007/s12036-025-10064-w},
       adsurl = {https://ui.adsabs.harvard.edu/abs/2025JApA...46...51N}
}

@ARTICLE{2015AdSpR..55.2745S,
       author = {{Schrijver}, Carolus J. and {Kauristie}, Kirsti and {Aylward}, Alan D. and {Denardini}, Clezio M. and {Gibson}, Sarah E. and {Glover}, Alexi and {Gopalswamy}, Nat and {Grande}, Manuel and {Hapgood}, Mike and {Heynderickx}, Daniel and {Jakowski}, Norbert and {Kalegaev}, Vladimir V. and {Lapenta}, Giovanni and {Linker}, Jon A. and {Liu}, Siqing and {Mandrini}, Cristina H. and {Mann}, Ian R. and {Nagatsuma}, Tsutomu and {Nandy}, Dibyendu and {Obara}, Takahiro and {Paul O'Brien}, T. and {Onsager}, Terrance and {Opgenoorth}, Hermann J. and {Terkildsen}, Michael and {Valladares}, Cesar E. and {Vilmer}, Nicole},
        title = "{Understanding space weather to shield society: A global road map for 2015-2025 commissioned by COSPAR and ILWS}",
      journal = {Advances in Space Research},
         year = 2015,
        month = jun,
       volume = {55},
       number = {12},
        pages = {2745-2807},
          doi = {10.1016/j.asr.2015.03.023},
archivePrefix = {arXiv},
       eprint = {1503.06135},
 primaryClass = {physics.space-ph},
       adsurl = {https://ui.adsabs.harvard.edu/abs/2015AdSpR..55.2745S}
}

@article{ishii2024pathways,
  title={Pathways to global coordination in space weather: International organizations, initiatives, and space agencies},
  author={Ishii, Mamoru and Costa, Joaquim Eduardo Rezende and Kuznetsova, Maria M and Andries, Jesse and Gopalswamy, Natchimuthuk and Belehaki, Anna and Alfonsi, Lucilla and Shiokawa, Kazuo and Stanislawska, Iwona and Bingham, Suzy and {International Space Weather Community}},
  journal={Advances in Space Research},
  year={2024},
  publisher={Elsevier}
}

@book{national2024next,
  title={The next decade of discovery in solar and space physics: Exploring and safeguarding humanity's home in space},
  author={{National Academies of Sciences, Engineering, and Medicine}},
  year={2024}
}

@ARTICLE{Sanchita_2024,
       author = {{Pal}, Sanchita and {G. dos Santos}, Luiz F. and {Weiss}, Andreas J. and {Narock}, Thomas and {Narock}, Ayris and {Nieves-Chinchilla}, Teresa and {Jian}, Lan K. and {Good}, Simon W.},
        title = "{Automatic Detection of Large-scale Flux Ropes and Their Geoeffectiveness with a Machine-learning Approach}",
      journal = {The Astrophysical Journal},
         year = 2024,
        month = sep,
       volume = {972},
       number = {1},
          eid = {94},
        pages = {94},
          doi = {10.3847/1538-4357/ad54c3},
archivePrefix = {arXiv},
       eprint = {2406.07798},
 primaryClass = {astro-ph.SR},
       adsurl = {https://ui.adsabs.harvard.edu/abs/2024ApJ...972...94P}
}

@ARTICLE{Grandin_2019,
       author = {{Grandin}, Maxime and {Aikio}, Anita T. and {Kozlovsky}, Alexander},
        title = "{Properties and Geoeffectiveness of Solar Wind High-Speed Streams and Stream Interaction Regions During Solar Cycles 23 and 24}",
      journal = {Journal of Geophysical Research (Space Physics)},
         year = 2019,
        month = jun,
       volume = {124},
       number = {6},
        pages = {3871-3892},
          doi = {10.1029/2018JA026396},
archivePrefix = {arXiv},
       eprint = {2006.06302},
 primaryClass = {physics.space-ph},
       adsurl = {https://ui.adsabs.harvard.edu/abs/2019JGRA..124.3871G}
}

@ARTICLE{Vysakh_2023,
       author = {{Vysakh}, P.~A. and {Mayank}, Prateek},
        title = "{Solar Flare Prediction and Feature Selection Using a Light-Gradient-Boosting Machine Algorithm}",
      journal = {Solar Physics},
         year = 2023,
        month = nov,
       volume = {298},
       number = {11},
          eid = {137},
        pages = {137},
          doi = {10.1007/s11207-023-02223-5},
archivePrefix = {arXiv},
       eprint = {2310.19332},
 primaryClass = {astro-ph.SR},
       adsurl = {https://ui.adsabs.harvard.edu/abs/2023SoPh..298..137V}
}

@ARTICLE{Bobra_2015,
       author = {{Bobra}, M. G. and {Couvidat}, S.},
        title = "{Solar Flare Prediction Using SDO/HMI Vector Magnetic Field Data with a Machine-learning Algorithm}",
      journal = {The Astrophysical Journal},
         year = 2015,
        month = jan,
       volume = {798},
       number = {2},
          eid = {135},
        pages = {135},
          doi = {10.1088/0004-637X/798/2/135},
archivePrefix = {arXiv},
       eprint = {1411.1405},
 primaryClass = {astro-ph.SR},
       adsurl = {https://ui.adsabs.harvard.edu/abs/2015ApJ...798..135B}
}

@ARTICLE{Mayank_2025,
       author = {{Mayank}, Prateek and {Camporeale}, Enrico and {Shrivastav}, Arpit K. and {Berger}, Thomas E. and {Arge}, Charles N.},
        title = "{Neural Enhancement of the Traditional Wang--Sheeley--Arge Solar Wind Relation}",
      journal = {The Astrophysical Journal Letters},
         year = 2025,
        month = nov,
       volume = {994},
       number = {1},
          eid = {L5},
        pages = {L5},
          doi = {10.3847/2041-8213/ae1735},
archivePrefix = {arXiv},
       eprint = {2509.06181},
 primaryClass = {astro-ph.SR},
       adsurl = {https://ui.adsabs.harvard.edu/abs/2025ApJ...994L...5M}
}

@ARTICLE{Enrico_2017,
       author = {{Camporeale}, Enrico and {Car{\`e}}, Algo and {Borovsky}, Joseph E.},
        title = "{Classification of Solar Wind With Machine Learning}",
      journal = {Journal of Geophysical Research (Space Physics)},
         year = 2017,
        month = nov,
       volume = {122},
       number = {11},
        pages = {10,910-10,920},
          doi = {10.1002/2017JA024383},
archivePrefix = {arXiv},
       eprint = {1710.02313},
 primaryClass = {physics.space-ph},
       adsurl = {https://ui.adsabs.harvard.edu/abs/2017JGRA..12210910C}
}

@ARTICLE{Andong_2023,
       author = {{Hu}, A. and {Camporeale}, E. and {Swiger}, B.},
        title = "{Multi-Hour-Ahead Dst Index Prediction Using Multi-Fidelity Boosted Neural Networks}",
      journal = {Space Weather},
         year = 2023,
        month = apr,
       volume = {21},
       number = {4},
          eid = {e2022SW003286},
        pages = {e2022SW003286},
          doi = {10.1029/2022SW003286},
archivePrefix = {arXiv},
       eprint = {2209.12571},
 primaryClass = {physics.space-ph},
       adsurl = {https://ui.adsabs.harvard.edu/abs/2023SpWea..2103286H}
}

@ARTICLE{Yuri_2019,
       author = {{Shprits}, Yuri Y. and {Vasile}, Ruggero and {Zhelavskaya}, Irina S.},
        title = "{Nowcasting and Predicting the Kp Index Using Historical Values and Real-Time Observations}",
      journal = {Space Weather},
         year = 2019,
        month = aug,
       volume = {17},
       number = {8},
        pages = {1219-1229},
          doi = {10.1029/2018SW002141},
       adsurl = {https://ui.adsabs.harvard.edu/abs/2019SpWea..17.1219S}
}

@ARTICLE{Hannah_2026,
       author = {{R{\"u}disser}, Hannah T. and {Nguyen}, Gautier and {Le Lou{\"e}dec}, Justin and {Davies}, Emma E. and {M{\"o}stl}, Christian},
        title = "{ARCANE--Early Detection of Interplanetary Coronal Mass Ejections}",
      journal = {Space Weather},
         year = 2026,
        month = feb,
       volume = {24},
       number = {2},
          eid = {e2025SW004537},
        pages = {e2025SW004537},
          doi = {10.1029/2025SW004537},
archivePrefix = {arXiv},
       eprint = {2505.09365},
 primaryClass = {physics.space-ph},
       adsurl = {https://ui.adsabs.harvard.edu/abs/2026SpWea..2404537R}
}

@ARTICLE{Sundararajan_2017,
       author = {{Sundararajan}, Mukund and {Taly}, Ankur and {Yan}, Qiqi},
        title = "{Axiomatic Attribution for Deep Networks}",
      journal = {arXiv e-prints},
         year = 2017,
        month = mar,
          eid = {arXiv:1703.01365},
        pages = {arXiv:1703.01365},
          doi = {10.48550/arXiv.1703.01365},
archivePrefix = {arXiv},
       eprint = {1703.01365},
 primaryClass = {cs.LG},
       adsurl = {https://ui.adsabs.harvard.edu/abs/2017arXiv170301365S}
}

@ARTICLE{Yogesh2026,
       author = {{Yogesh} and {Ofman}, Leon and {Klein}, Kristopher G. and {Niranjana} and {Martinovi{\'c}}, Mihailo and {Howes}, Gregory G. and {Mostafavi}, Parisa and {Boardsen}, Scott A. and {Sadykov}, Viacheslav M. and {Pal}, Sanchita and et al.},
        title = "{Solar Wind Heating near the Sun: A Radial Evolution Approach}",
      journal = {The Astrophysical Journal},
         year = 2026,
        month = mar,
       volume = {999},
       number = {2},
          eid = {225},
        pages = {225},
          doi = {10.3847/1538-4357/ae4582},
archivePrefix = {arXiv},
       eprint = {2602.10275},
 primaryClass = {astro-ph.SR},
       adsurl = {https://ui.adsabs.harvard.edu/abs/2026ApJ...999..225Y}
}

@article{riley_2011_hux,
  author = {Riley, Pete and Lionello, Roberto},
  month = {05},
  pages = {575-592},
  title = {Mapping Solar Wind Streams from the Sun to 1 AU: A Comparison of Techniques},
  doi = {10.1007/s11207-011-9766-x},
  urldate = {2023-05-29},
  volume = {270},
  year = {2011},
  journal = {Solar Physics}
}

@ARTICLE{Mayank_2025_swasti,
       author = {{Mayank}, Prateek and {J. Athalathil}, Jithu and {Nandy}, Sirsha and {Vaidya}, Bhargav and {Navanit}, A.~V. and {Paul}, Arghyadeep},
        title = "{SWASTi: A physics-based modelling toolkit for space weather}",
      journal = {Journal of Astrophysics and Astronomy},
         year = 2025,
        month = oct,
       volume = {46},
       number = {2},
          eid = {80},
        pages = {80},
          doi = {10.1007/s12036-025-10107-2},
       adsurl = {https://ui.adsabs.harvard.edu/abs/2025JApA...46...80M}
}

@ARTICLE{2014ApJ...782...81V,
       author = {{van der Holst}, B. and {Sokolov}, I.~V. and {Meng}, X. and {Jin}, M. and {Manchester}, IV, W.~B. and {T{\'o}th}, G. and {Gombosi}, T.~I.},
        title = "{Alfv{\'e}n Wave Solar Model (AWSoM): Coronal Heating}",
      journal = {The Astrophysical Journal},
         year = 2014,
        month = feb,
       volume = {782},
       number = {2},
          eid = {81},
        pages = {81},
          doi = {10.1088/0004-637X/782/2/81},
archivePrefix = {arXiv},
       eprint = {1311.4093},
 primaryClass = {astro-ph.SR},
       adsurl = {https://ui.adsabs.harvard.edu/abs/2014ApJ...782...81V}
}

@article{platt1999,
  title={Probabilistic outputs for support vector machines and comparisons to regularized likelihood methods},
  author={Platt, John},
  journal={Advances in large margin classifiers},
  volume={10},
  number={3},
  pages={61--74},
  year={1999},
  publisher={Cambridge, MA}
}

@ARTICLE{Gosling1978,
       author = {{Gosling}, J.~T. and {Asbridge}, J.~R. and {Bame}, S.~J. and {Feldman}, W.~C.},
        title = "{Solar wind stream interfaces}",
      journal = {Journal of Geophysical Research},
         year = 1978,
        month = apr,
       volume = {83},
       number = {A4},
        pages = {1401-1412},
          doi = {10.1029/JA083iA04p01401},
       adsurl = {https://ui.adsabs.harvard.edu/abs/1978JGR....83.1401G}
}

@ARTICLE{Rout2017,
       author = {{Rout}, Diptiranjan and {Chakrabarty}, D. and {Janardhan}, P. and {Sekar}, R. and {Maniya}, Vrunda and {Pandey}, Kuldeep},
        title = "{Solar wind flow angle and geoeffectiveness of corotating interaction regions: First results}",
      journal = {Geophysical Research Letters},
         year = 2017,
        month = may,
       volume = {44},
       number = {10},
        pages = {4532-4539},
          doi = {10.1002/2017GL073038},
       adsurl = {https://ui.adsabs.harvard.edu/abs/2017GeoRL..44.4532R}
}

@ARTICLE{AdamW_losh_hutter,
       author = {{Loshchilov}, Ilya and {Hutter}, Frank},
        title = "{Decoupled Weight Decay Regularization}",
      journal = {arXiv e-prints},
         year = 2017,
        month = nov,
          eid = {arXiv:1711.05101},
        pages = {arXiv:1711.05101},
          doi = {10.48550/arXiv.1711.05101},
archivePrefix = {arXiv},
       eprint = {1711.05101},
 primaryClass = {cs.LG},
       adsurl = {https://ui.adsabs.harvard.edu/abs/2017arXiv171105101L}
}

@ARTICLE{sunrunner3D,
       author = {{Gonz{\'a}lez-Avil{\'e}s}, Jos{\'e} Juan and {Riley}, Pete and {Ben-Nun}, Michal and {Mayank}, Prateek and {Vaidya}, Bhargav},
        title = "{Using sunRunner3D to interpret the global structure of the heliosphere from in situ measurements}",
      journal = {Journal of Space Weather and Space Climate},
         year = 2024,
        month = aug,
       volume = {14},
          eid = {12},
        pages = {12},
          doi = {10.1051/swsc/2024014},
archivePrefix = {arXiv},
       eprint = {2405.00174},
 primaryClass = {astro-ph.SR},
       adsurl = {https://ui.adsabs.harvard.edu/abs/2024JSWSC..14...12G}
}
%


%
%
%
%
%

\end{document}